\documentclass[nofootinbib,amsfonts,prd,aps]{revtex4}
\usepackage{subfig}
\usepackage{amsmath,amssymb,amsfonts,amsthm,mathrsfs}
\usepackage{stackrel}
\usepackage{graphicx}
\usepackage{nameref}
\DeclareMathOperator{\sech}{sech}
\begin{document}

\title{Hawking radiation from an evaporating black hole via Bogoliubov transformations}

\author{Rodrigo Eyheralde}
\affiliation {
Instituto de F\'{\i}sica, Facultad de Ciencias, 
Igu\'a 4225, esq. Mataojo, 11400 Montevideo, Uruguay.}

\begin{abstract}
  We study Hawking radiation on a Vaidya space-time with a gravitational collapse followed by evaporation. The collapsing body is a null thin-shell and the evaporation is induced by a negative energy collapsing null-shell. This mimics the back-reaction to the Hawking radiation. Using Hawking's original method of Bogoliubov transformations we characterize the radiated spectrum as dominated by a thermal emission with an increasing effective temperature. We compute this time dependent temperature and find numerical agreement with results obtained by other techniques. The known divergences at the evaporation time are explained by the divergent nature of the effective temperature. As a consistency check, we re-derived the results from a zero mass limit of a remnant $BH$ scenario.
\end{abstract}

\maketitle

\section{Introduction}

Hawking radiation plays a key role in the discussion of the tension between General Relativity and Quantum Mechanics. It is required for the interpretation of the Bekeinstein-Hawking formula as a thermodynamic entropy but also presents a mechanism for Black Holes (BHs) to emit energy and evaporate. This leads to information paradoxes \citep{marolf2017,Unruh:2017uaw} which are subject of intense debate. The consensus is that the description of the evaporation process, including the resolution of the paradox, requires a theory of Quantum Gravity. For example, in a recent series of papers \citep{Penington:2019npb,Almheiri:2019psf,Almheiri:2020cfm} the authors propose an entropy computation without paradoxical behavior (following Page curve \citep{Page:1993wv,Page:2013dx}) and which explicitly relies on physics beyond classical gravity. Even if the problem of the Page curve is solved, a complete picture of the evaporation would need to include what happens when the BH reaches Plank size and in particular, how the Hawking effect is modified there.

As a first step, in this paper we revisit Hawking's computation applied to such extreme scenario using a Vaidya metric as a toy model for evaporation. In such model the BH is created by collapsing massless particles and the evaporation is not driven by the back-reaction itself but by the collapse of negative energy massless particles. This kind of evaporating black hole (EBH) models have been considered before \citep{Hiscock:1980ze,Frolov:2016gwl,Bianchi:2014bma} and some aspects of the radiation such as luminosity and entropy flux has been computed, under spherical symmetry assumption, with conformal anomaly techniques.

To our knowledge Hawking's original method of Bogoliubov transformations has not been directly applied to EBH models. Instead, several other methods have been developed, most of which focus in the computation of the renormalized (two dimensional) energy momentum tensor \citep{Davies:1976ei,Fabbri:2004yy}. In fact it has been argued \citep{Hiscock:1980ze} that Hawking's method is not suited for such scenarios  where the space-time is not globally hyperbolic. Here we argue that Bogoliubov transformations can be applied if we restrict the computation to the appropriate outgoing radiative modes. This provides a complete description of the radiation profile and we find that it is dominated by a thermal emission with a growing Hawking temperature as is expected to happen in the early stages of a real evaporation of a macroscopic BH.

In section II we obtain an expression for the density matrix of a massless scalar field via Bogoliubov transformations and obtain expressions for some relevant observables built from it. The analogous computation for a system with collapse but without evaporation is included in Appendix A and it provides us a comparison point for the EBH case. This appendix may be of interest on its own because it presents what we believe to be a new method to compute the Hawking profile from the Bogoliubov coefficients. In section III we analyze the results, present the thermal profile and show the numerical agreement with known techniques from the literature. Finally, in section IV we compare our computation with a remnant scenario in the limit where the mass of the remnant BH goes to zero. We re-obtain the same result in a scenario where the Cauchy problem for the scalar field is better defined and confirm the interpretation of the effective temperature described in section III. We also comment on the relation of our model to similar metrics known as regular Black Holes, where a Planck scale modification is introduced to the metric. These are examples of globally hyperbolic space-times where the (dramatic) departure from our results are detected by far away observers close before the end of the emission due to the complete evaporation.

\section{Particle emission}
\label{sec:emision}
\begin{figure}
\includegraphics[height=11cm]{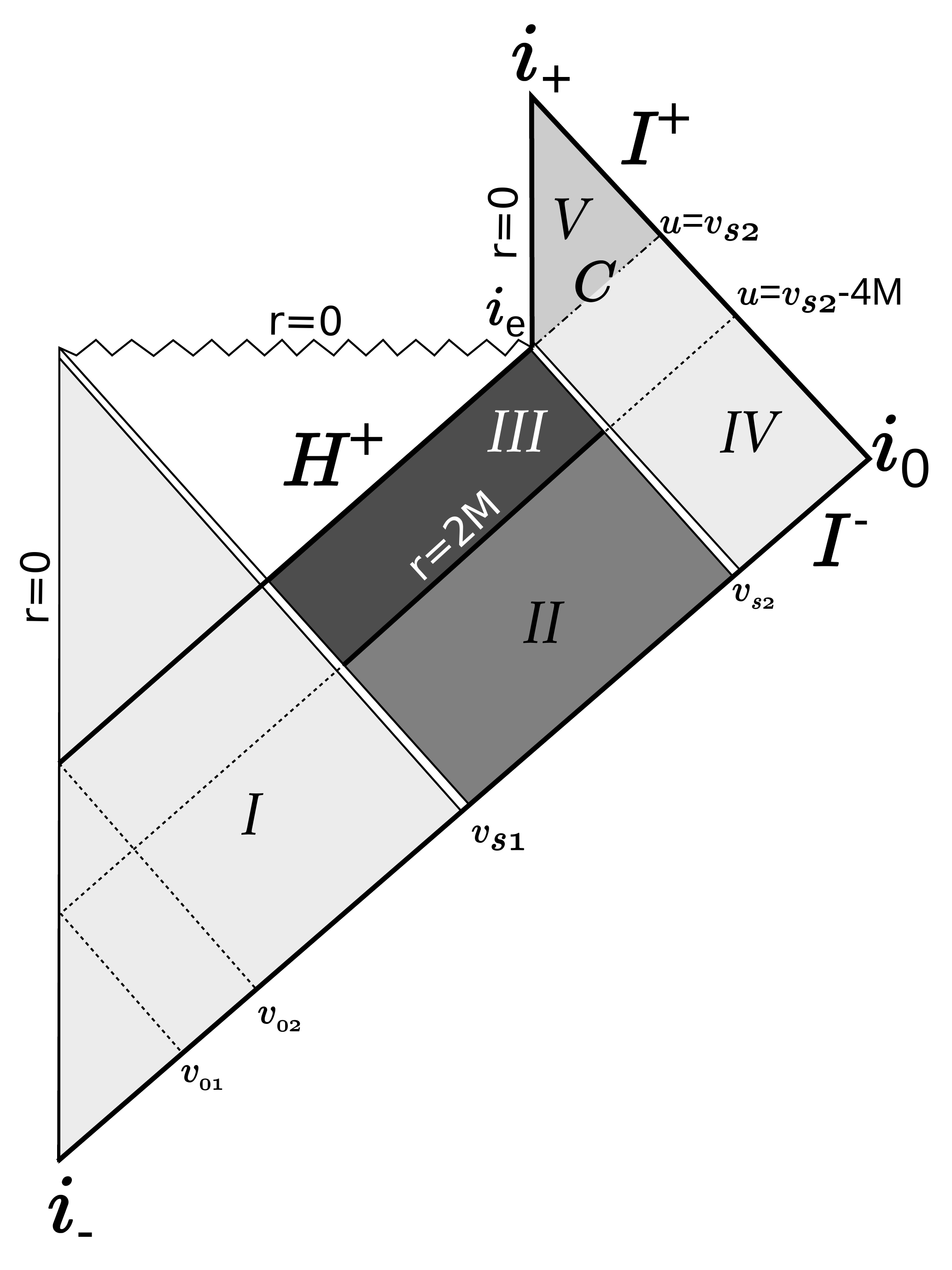}
\caption{Penrose-Carter diagram for metric (\ref{eq:metrica}). $v_{s1}$ and $v_{s2}$ are the null coordinates at $I^-$ of the first and second shell, respectively. Light rays sent in before $v_{02}$ or after $v_{s2}$ reach $r=0$ and bounce back to $I^+$, whereas rays sent in between get trapped in the BH interior. Highlighted regions are Minkowski regions ($I$, $IV \& V$), exterior Schwarzschild ($II$) and the trapped surfaces region ($III$) between the apparent and even horizons.}
\label{fig:sandwich}
\end{figure}

As a simple model of collapse and evaporation we consider the Vaidya metric given by the line element
\begin{equation}
ds^{2}=-\left(1-\frac{2M(v)}{r}\right)dv^{2}+2dvdr+r^{2}d\Omega^{2},\label{eq:metrica}
\end{equation}
with 
\begin{equation}
M(v)=M\left[\theta(v-v_{s1})-\theta(v-v_{s2})\right].
\end{equation}
The parameter $v_{s1}$ represents the position of a collapsing shell (using ingoing Eddington--Finkelstein/advanced time coordinates at $I^-$) with energy $M$ whereas $v_{s2}>v_{s1}$ represents the position of a second shell with energy $-M$. This geometry includes an apparent horizon at $r=2M$ between the arrival of the first and second shell. There is also an event horizon ($H^+$) formed by the past of the outward null rays that arrive at the singularity exactly when the second shell does. They meet at the \textit{evaporation point} $i_e$ and from there a Cauchy horizon ($C$) expands towards $I^+$. We will refer as \textit{evaporation time} to the retarded time in $I^+$ of the null rays coming from $i_e$.  In this \textit{``sandwich''} \citep{Frolov:2016gwl} space-time (see Figure \ref{fig:sandwich}) we will refer to the Minkowski regions as $I$ (inside the first shell), $IV$ (outside the second shell) and $V$ (after evaporation); $II$ will be the Schwarzschild region outside the apparent horizon and $III$ the region of trapped surfaces outside $H^+$.

On the above metric we would like to study Hawking radiation corresponding to a massless scalar field using Hawking's original method of Bogoliubov transformations \citep{h74}. In this method two quantizations of the fields in the fixed background metric (\ref{eq:metrica}) are compared. One is the \textit{in} construction, which is the quantization associated with the preferred vacuum for observers in the far past and whose one particle Hilbert space is spanned by positive frequency spherical modes 
\begin{equation}
\psi^{in}_{lm\omega'}(r,v,\theta,\phi)=\frac{\phi_{l\omega'}(v,r)}{ r\sqrt{4\pi\omega'}}Y_{l}^m(\theta,\phi),\quad \omega'>0,\label{eq:in}
\end{equation}
which are given by the asymptotic form
\begin{equation}
\phi_{l\omega'}(v,r)\underset{r\to\infty}{\longrightarrow}e^{-i\omega'v},
\end{equation}
at $I^-$. The other quantization is the \textit{out-int} construction corresponding to the preferred vacuum for observers in the far future (or at the Horizon) and whose one particle Hilbert space is spanned by positive frequency \textit{out} modes 
\begin{equation}
\chi^{out}_{lm\omega}(r,u,\theta,\phi)=\frac{\chi_{l\omega}(u,r)}{ r\sqrt{4\pi\omega}}Y_{l}^m(\theta,\phi),\quad \omega>0.
\label{eq:out}
\end{equation}
with asymptotic form
\begin{equation}
\chi_{l\omega}(u,r)\underset{r\to\infty}{\longrightarrow}e^{-i\omega u},
\end{equation}
at $I^+$ which vanish at the event horizon ($H^+$) plus \textit{int} modes that vanish at $I^+$ and register at $H^+$. Here $u$ is the outgoing Eddington-Finkelstein/retarded time coordinate. We argue that this modes are well defined across the Cauchy horizon ($C$) and span a subspace of radiative solutions outside $H^+$ which are enough to describe the Hawking effect. We present the argument in subsection \ref{sec:bogol} and provided detailed calculations in appendix \ref{sec:KG}.

\subsection{Near-Horizon approximation}

The Hawking effect only depends on the relation between the creation and annihilation operators associated to the \textit{out} and the \textit{in} modes so the \textit{int} modes can be traced out. This relation is given by the Bogoliubov coefficients \citep{h74},
\begin{eqnarray}
\alpha_{lm\omega,l'm'\omega'}&=&\left\langle \chi^{out}_{lm\omega},\psi^{in}_{l'm'\omega'}\right\rangle,\\
\beta_{lm\omega,l'm'\omega'}&=&-\left\langle \chi^{out}_{lm\omega},(\psi^{in}_{l'm'\omega'})^{*}\right\rangle
\end{eqnarray}
written as Klein Gordon (K-G) product of the modes which allows to compute them in any Cauchy surface. In particular the $\beta$ coefficients encode the information about the particle content of the \textit{in} vacuum seen by observes in the \textit{out} region. There, the state is represented by the density matrix \citep{h74}
\begin{equation}
\rho_{l_1 m_1\omega_1,l_2 m_2\omega_2}=\sum_{l',m'}{\int}_0^\infty d\omega'\beta_{l_1 m_1\omega_1,l'm'\omega'}\beta_{l_2 m_2\omega_2,l'm'\omega'}^{*},
\end{equation}
whose diagonal is the flux of emitted particles $N_{lm\omega}=\rho_{lm\omega,lm\omega}$.

Since the K-G product is invariant under change of Cauchy surface we chose to compute it at $I^-$, which requires the projection of the \textit{out} modes to the past. This is a difficult calculation because this modes are governed by the wave equation
\begin{equation}
\left[-4\frac{\partial^2}{\partial u\partial v}-V_l(r,M(v))\right]\phi_{l\omega'}=0\label{eq:KG}
\end{equation}
where $\pm V_l(r,M)=\pm\left(1-\frac{2M}{r}\right)\left[\frac{l(l+1)}{r^2}+\frac{2M}{r^3}\right]$ acts as a the ``centrifugal potential'' with $+$ for region $II$ and $-$ for region $III$\footnote{The sign arises because the $u$ Eddington-Finkelstein coordinate grows as one approaches the apparent horizon in both regions $II$ and $III$.}. However in the \textit{near-horizon approximation} the potential is neglected and geometric optics can be used to trace wave solutions as they propagate between Cauchy surfaces. In particular, the trajectory of null rays control the phase of the waves. From here on we will consider such assumption but it should be pointed out that it is valid only near the apparent horizon (interface of regions $II$ and $III$), from where it draws its name, also for $l=0$ modes in the Minkowski regions ($I$, $IV$ and $V$) and at early times in region $II$. On the other hand, it is a bad approximation near $i_e$ in region $III$, where the potential diverges. Correction the this approximation are expected to appear in the form of frequency dependent gray body factors as in the original Hawking calculation \citep{h74}. In what follow we reduce our study to the \textit{s-modes} ($l=0$) in order for the approximation to be exact in the Minkowski regions $I,IV$ and $V$. This extra assumption also reduces the problem to an effective two dimensional problem and introduces a conformal symmetry in the wave equation allowing the use of conformal anomaly techniques to compute some observables of the quantum field, as we discuss in subsection \ref{sec:conformal}. This provides a comparison point to check the consistency of our calculation and we will make extensive use of it in the next section.

\begin{figure}
\includegraphics[height=7cm]{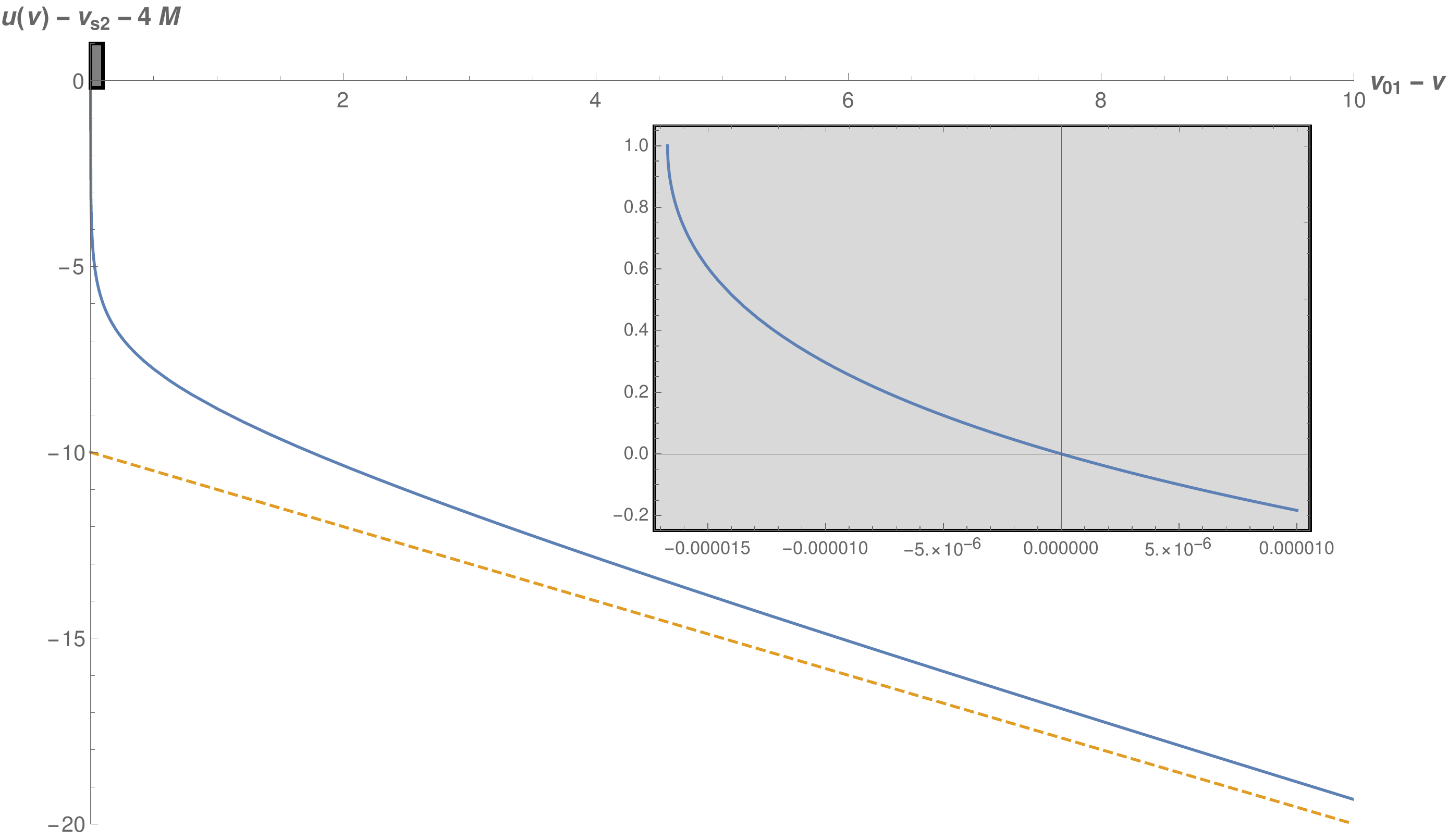}
\caption{Retarded time coordinate $U(v)$ for the outgoing rays that left $I^-$ before the formation of the event horizon ($v<v_{02}$ in region $I$). Highlighted is the region of rays coming out of region $III$ ($v_{s2}-4M<U(v)<v_{s2}$). The lower bound (dashed line) is the asymptotic past behavior ($U(v)\to v$) where rays  propagate far from the BH. Everything is presented in units of $4M$ and plotted for $\Delta v_s = 10$.}
\label{fig:u}
\end{figure}

Applying geometric optics to trace the modes requires the description the light ray trajectories. Geometry (\ref{eq:metrica}) has three families of radial light rays leaving $I^{-}$ with coordinate $v$. For 
\begin{equation}
v<v_{02}=v_{s1}-4M-4M\textrm{W}\left[-\exp\left(-1-\Delta v_s\right)\right]
\end{equation}
the rays bounce through $r=0$ and arrive at $I^{+}$ with outgoing Eddington--Finkelstein coordinate (see Figure \ref{fig:u})
\begin{equation}
U(v)=v_{s2}-4M-4M\textrm{W}\left[\frac{v_{s1}-v-4M}{4M}\exp\left(\frac{v_{s2}-v-4M}{4M}\right)\right]\label{eq:u(v)}
\end{equation}
where $\rm{W}$ is the principal branch of the Lambert W function\footnote{The Lambert $\rm{W}$ function is defined as the inverse of $f(x)=x\exp(x)$. It is a multivalued function with branches $W_{k\in\mathbb{Z}}$. In this paper we refer always to the principal branch ($\rm{W}\equiv W_0$) which is the analytic extension of the real $f^{-1}$ in the domain $(-1/e,+\infty)$ and range $(-1,+\infty)$. The real inverse with domain $(-1/e,0)$ and range $(-\infty,-1)$ corresponds to the branch $W_{-1}$.} and $\Delta v_s=\frac{v_{s2}-v_{s1}}{4M}$. 
A second family is formed by rays with $v_{02}<v<v_{s2}$ which fall into the singularity and a third family of rays with $v>v_{s2}$ move through flat space-time, reach $r=0$ after evaporation and bounce to $I^+$ with $U(v)=v$. The ray at $v=v_{02}$ defines the position of the event and Cauchy horizons and the one at $v=v_{01}=v_{s1}-4M$ identify the position of the apparent horizon.

\subsection{Bogoliubov coefficients}\label{sec:bogol}
The near-horizon approximation together with the imposition of regularity at $r=0$ and continuity at the shells as Israel matching conditions \citep{Israel:1966rt} or by Penrose copy-paste formalism (see references in \citep{Manzano:2021auk}) allow us to project radiative modes throughout the exterior of the BH horizon $H^+$. This is key for the computation of Bogoliubov coefficients. Detailed calculations are provided in appendix \ref{sec:KG} and the relevant results are shown in what follows. 

The \textit{s-modes} in the \textit{out} construction defined by (\ref{eq:out}) acquire the form 
\begin{equation}
\chi_{00\omega}^{out}(r,v,\theta,\phi)\equiv\chi_{\omega}^{out}(r,v)=\frac{1}{4\pi r\sqrt{\omega}}\times\left\{ \begin{array}{lcl}
e^{-i\omega u}-e^{-i\omega v},\qquad v>v_{s2}\quad (IV \& V)\\
e^{-i\omega \hat{U}(\bar{u},v_{s2})}-e^{-i\omega v_{s2}},\qquad v_{s2}>v>v_{s1}\quad (II \& III)\\
e^{-i\omega U(u)}-e^{-i\omega v_{s2}},\qquad v_{s1}>v>v_{02}\quad (I)\\
e^{-i\omega U(u)}-e^{-i\omega U(v)},\qquad v<v_{02}\quad (I)
\end{array}\right.
\end{equation}
in the bulk, where $\bar{u}=v-2r-4M\ln\left|\frac{r-2M}{2M_0}\right|$ is the outgoing Eddington-Finkelstein coordinate in regions $II$ and $III$ while $u=v-2r$ is the retarded time in regions $I$ and $IV\&V$. The auxiliary function $\hat{U}$ is defined by (\ref{eq:Uhat}). On the other hand, any \textit{int} spherically symmetric mode $\chi^{int}(r,v)$ given by
\begin{equation}
r\chi^{int}(r,v)\to F(v)
\end{equation}
at $H^+$ and vanishing at $I^+$ also vanishes in regions $I,IV\&V$ while keeping the form $\chi^{int}(v)=F(v)/r$ in regions $II$ and $III$ provided it fulfills the continuity condition $F(v_{02})=F(v_{s2})=0$.
Notice that the \textit{out} modes reduce to  
\begin{equation}
\chi_{\omega}(v)=\frac{1}{4\pi r\sqrt{\omega}}\left(e^{-i\omega v_{s2}}-e^{-i\omega v}\right)
\end{equation}
at the Cauchy horizon $C$ and this guaranties its continuity across $H^+\cup C$. Considering the Cauchy surface $$\Sigma=H^+\cup C\cup I^+|_{u<v_{s2}}$$
the previous calculation shows that \textit{out} modes register at $C\cup I^+|_{u<v_{s2}}$ while \textit{int} modes register only at $H^+$. Therefore they are orthogonal with respect to the K-G product and this allows to trace out the \textit{int} modes to compute the density matrix of the radiation as in the original Hawking calculation.

To confirm the consistency of the method we can project the \textit{in} modes (\ref{eq:in}) to the bulk (outside $H^+$) under the same approximations and obtain
\begin{equation}
\psi_{\omega'}(r,v)=\frac{1}{4\pi r\sqrt{\omega'}}\times\left\{ \begin{array}{lcl}
e^{-i\omega' v}-e^{-i\omega' u},\qquad v<v_{s1}\quad (I)\\
e^{-i\omega' v}-e^{-i\omega' \hat{U}(\bar{u},v_{s1})},\qquad v_{s2}>v>v_{s1}\quad (II \& III)\\
e^{-i\omega' v}-e^{-i\omega' \tilde{U}(u)},\qquad v>v_{s2}\quad (IV)\\
\end{array}\right.
\end{equation}
where $\tilde{U}(u)$ is analogous to $U(v)$ and is given by (\ref{eq:Utilde}). Again, the restriction to $H^+\cup C$ gives a continuous function even in the evaporation point $i_e$. This construction shows the computation of the K-G product between an \textit{out} and an \textit{in} mode is well defined at $\Sigma$ and then it can be projected back to $I^-$. The projection of \textit{in} modes to the future of $C$ is ill defined and is discussed in appendix \ref{sec:KG}, but it doesn't affect the computation of the relevant Bogoliubov coefficients that we present in what follows.

The projection of \textit{out} s-modes to $I^-$ allow us to compute the Bogoliubov coefficient
\begin{equation}
\beta_{\omega\omega'}=\frac{1}{2\pi}\sqrt{\frac{\omega'}{\omega}}\left[
{\int_{-\infty}^{v_{02}}}dv\left(e^{-i\omega v_{s2}}-e^{-i\omega U(v)}\right)e^{- i\omega'v}+{\int_{v_{s2}}^{+\infty}}dv\left(e^{-i\omega v_{s2}}-e^{-i\omega v}\right)e^{- i\omega'v}\right].\label{eq:beta}
\end{equation}
The divergent nature of such integrals comes from the fact that we are considering delta normalized modes. For this reason it is convenient to consider wave-packets localized in both frequency and time. For example the set $\left\lbrace\chi_{n\omega_{j}}\right\rbrace$ given by
\begin{equation}
\chi^{out}_{n\omega_{j}}=\frac{1}{\sqrt{\epsilon}}{\int_{j\epsilon}^{\left(j+1\right)\epsilon}}d\omega e^{u_n\omega i}\chi_{\omega}^{out},\label{eq:paquete_def}
\end{equation}
is a countable orthonormal basis of packets centered in time $u_n=\frac{2\pi n}{\epsilon}$, and in frequency $\omega_{j}=\left(j+\frac{1}{2}\right)\epsilon$. Each wave-packet include a narrow set of frequencies and is heavily peaked around $u_n$ but it includes a \textit{tail} expanding the entire range of time $u$. This is an inescapable feature of this method and will be shown to be relevant in the calculation. 

Choosing $\epsilon<<\omega_j$ the $\beta$ coefficient for a wave packet can be approximated by
\begin{equation*}
\beta_{\omega_j\omega'}(u_n) \sim I_{v<v_{02}}+I_{v>v_{s2}}
\end{equation*}
where
\begin{eqnarray}
I_{v<v_{02}} & = &\frac{1}{2\pi}\sqrt{\frac{\omega'\epsilon}{\omega_j}}
{\int_{-\infty}^{v_{02}}}dv\left(e^{i\omega_j\left[u_n-v_{s2}\right]}{\rm{sinc}}\left[(u_n-v_{s2})\epsilon/2\right]-e^{i\omega_j\left[u_n-U(v)\right]}{\rm{sinc}}\left[(u_n-U(v))\epsilon/2\right]\right)e^{- i\omega'v},\label{eq:beta_relevant}\\
I_{v>v_{s2}} & = &\frac{1}{2\pi}\sqrt{\frac{\omega'\epsilon}{\omega_j}}{\int_{v_{s2}}^{+\infty}}dv\left(e^{i\omega_j\left[u_n-v_{s2}\right]}{\rm{sinc}}\left[(u_n-v_{s2})\epsilon/2\right]-e^{i\omega_j\left[u_n-v\right]}{\rm{sinc}}\left[(u_n-v)\epsilon/2\right]\right)e^{- i\omega'v}.\label{eq:beta_trivial}
\end{eqnarray}
The second contribution (\ref{eq:beta_trivial}) to the integral doesn't know about the collapse and falls rapidly with $n$ away from $u_n=v_{s2}$. To see this we can use the freedom in the definition of $u$ to choose $u_n-v_{s2}=\frac{2n\pi}{\epsilon}$ and the c.o.v. $y=(v-v_{s2})\epsilon/2$. Then the contribution becomes\footnote{Notice that the $\epsilon$ parameter in (\ref{eq:beta_trivial2}) can be absorbed in $\omega'$ since the observables of interest such as $N_{\omega_j}$ involve an extra integration in $\omega'>0$. }
\begin{eqnarray}
I_{v>v_{s2}} &=&\frac{e^{i(2j+1)n\pi-i\omega'v_{s2}}}{\pi\epsilon}\sqrt{\frac{\omega'}{j+1/2}}{\int_{0}^{+\infty}}dy\left({\rm{sinc}}\left[n\pi\right]-e^{-i(2j+1)y}{\rm{sinc}}\left[n\pi-y\right]\right)e^{- i\frac{2\omega'}{\epsilon}y}=\nonumber\\
&=&\frac{e^{-i\omega'u_n}}{2\pi\epsilon}\sqrt{\frac{\omega'}{j+1/2}}\left({\rm{Ei}}\left[2n\pi(j+\omega'/\epsilon)i\right]-{\rm{Ei}}\left[-2n\pi(j+1+\omega'/\epsilon)i\right]+\delta_{n0}\frac{\epsilon}{\omega'}\right)\label{eq:beta_trivial2}
\end{eqnarray}
where ${\rm{Ei}}$ is the Exponential integral function with the asymptotic behavior
\begin{equation}
{\rm{Ei}}(z)=\frac{e^z}{z}\left[\underset{k=0}{\overset{n}{\sum}}\frac{k!}{z^k}+O^\infty\left(z^{-n-1}\right)\right].
\end{equation}
From now on we neglect $I_{v>v_{s2}}$.

To evaluate the first contribution (\ref{eq:beta_relevant}) we use the integration variable
\begin{equation}
z\equiv\frac{v_{s1}-v-4M}{4M}\textrm{exp}\left(\frac{v_{s2}-v-4M}{4M}\right),
\end{equation}
with range $[-e^{-1},+\infty)$ whose inverse is 
\begin{equation}
v=v_{01}-4MW\left(ze^{-\Delta v_s}\right).
\end{equation}
Considering only this contribution to the Bogoliubov coefficient and choosing $u_n-v_{s2}=2n\pi/\epsilon$, the expression for the $\beta$ coefficient is
\small{\begin{equation}
\beta_{\omega_j\omega'}(n)\sim \frac{4Me^{-i4M\omega'v_{01}+i\omega_j\frac{2\pi n}{\epsilon}}}{2\pi}\sqrt{\frac{\omega'\epsilon}{\omega_j}}
\stackrel[-e^{-1}]{+\infty}{\int}\frac{dz}{z}\frac{W\left(ze^{-\Delta v_s}\right)e^{i4M\omega'W\left(ze^{-\Delta v_s}\right)}}{\left[1+W\left(ze^{-\Delta v_s}\right)\right]}\left[\delta_{n0}-e^{i4M\omega_j\left[1+W(z)\right]}{\rm{sinc}}\left(n\pi+\left[1+W(z)\right]2M\epsilon\right)\right].
\end{equation}}
From here on we can exclude the case $n=0$ since it correspond to radiation coming from the evaporation point $i_e$ and we don't expect the model to accurately describe particle emission there. The analytic extension of $W(z)$ has a branch cut in $z\in(-\infty,-e^{-1})$ and the integrand presents no other poles so it can be Wick-rotated\footnote{In \citep{Gao:2002kz} it is argued that Wick rotations is ambiguous in this kind of integral expressions if done for monochromatic spherical waves and can lead to erroneous results. Therefore the use of a wave packets basis is mandatory at this point of the calculation.} to the contour
\begin{equation}
z[t]=-e^{-1}+ie^t,\quad t\in\mathbb{R}.\label{eq:z_t}
\end{equation}
Finally, 
\small{\begin{equation}
\beta_{\omega_j\omega'}(n)= \frac{-i4Me^{-i4M\omega'v_{01}+i\omega_j\frac{2\pi n}{\epsilon}}}{2\pi}\sqrt{\frac{\omega'\epsilon}{\omega_j}}
\stackrel[-\infty]{+\infty}{\int}\frac{dte^t W\left(z[t]e^{-\Delta v_s}\right)e^{i4M\omega'W\left(z[t]e^{-\Delta v_s}\right)}e^{i4M\omega_j\left[1+W\left(z[t] \right)\right]}}{z[t]\left[1+W\left(z[t]e^{-\Delta v_s}\right)\right]}{\rm{sinc}}\left(n\pi+\left[1+W\left(z[t]\right)\right]2M\epsilon\right).\label{eq:Bogoliubov_final}
\end{equation}}
This coefficients are the building blocks for the observables we will compute in the rest of the section.
\subsection{Density matrix, particles flux and luminosity}
From Bogoliubov coefficients we can obtain the density matrix $\rho$ which, in principle, has the complete information of the mixed state corresponding to the Hawking radiation.

\begin{figure}
\centering
\subfloat[{Real part of $W(z[t])$ for $\Delta v_s=10$ with its asymptotic behavior at $t=\pm\infty$ (dashed lines).}]{\includegraphics[scale=0.5]{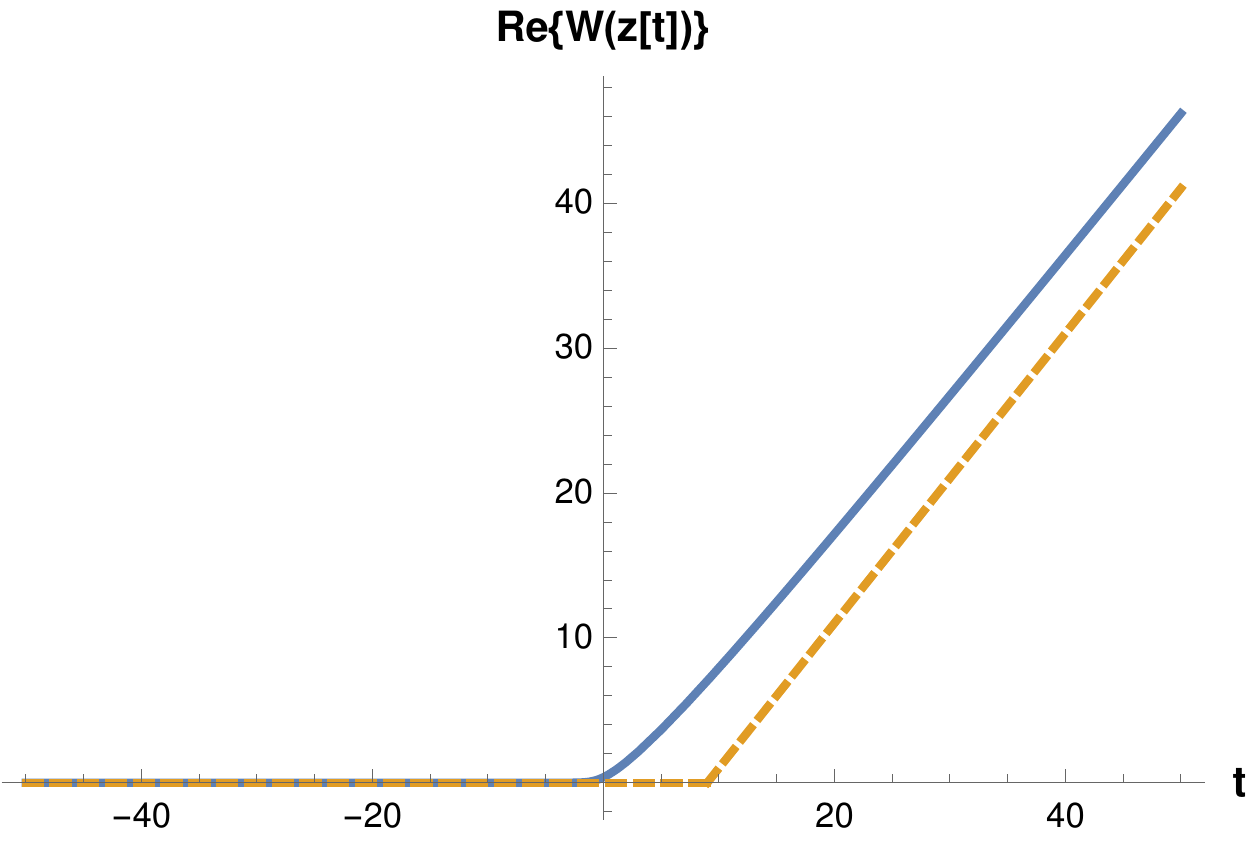}}
\qquad
\subfloat[{Imaginary part of $W(z[t])$ for $\Delta v_s=10$ with its asymptotic values ($0$ and $\frac{\pi}{2}$) given by the dashed lines.}]{\includegraphics[scale=0.5]{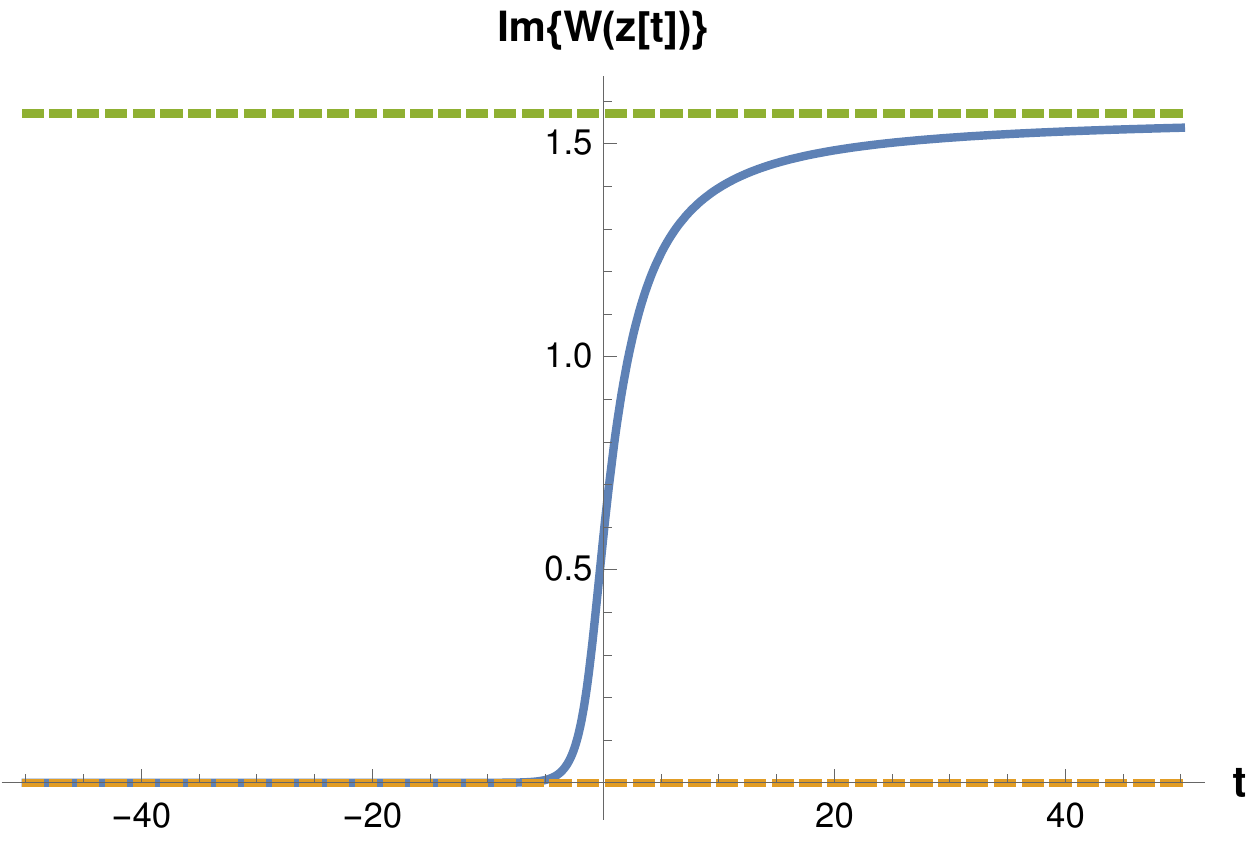}}\caption{Complex function resulting from the composition of the analytic extension of $W(z)$ with $z[t]$.}\label{fig:W}
\end{figure}

Since $W$ is the analytic extension of the Lambert function (therefore $W^*(z)=W(z^*)$) and $\textit{Im}\left[W\left(z[t]e^{-\Delta v_s}\right)\right]$ is positive for all $t$ (see Figure \ref{fig:W}), the $\omega'$ integral in
\[
\rho_{\omega_j\omega_k} ={\int}_0^\infty d\omega'\beta_{\omega_j\omega'}\beta_{\omega_k\omega'}^{*}
\]
can be computed inverting the order of integration. We get
\begin{eqnarray}
\rho_{\omega_j\omega_k}(n)&=&\frac{1}{2\pi^2\sqrt{(2j+1)(2k+1)}}\stackrel[-\infty]{+\infty}{\iint} dtd\bar{t}
G(t,\bar{t},\Delta v_s)e^{i4M\omega_j\left[1+W\left(z[t] \right)\right]-i4M\omega_k\left[1+W\left(z[\bar{t}]^*\right)\right]}\times\nonumber\\
&\times &\textrm{sinc}\left(n\pi+\left[1+W\left(z[t] \right)\right]2M\epsilon\right)\textrm{sinc}\left(n\pi+\left[1+W\left(z[\bar{t}]^*\right)\right]2M\epsilon\right).\label{eq:numero_varz}
\end{eqnarray}
where
\begin{equation}
G(t,\bar{t},\Delta v_s)=\frac{-1 e^te^{\bar{t}}W\left(z[t]e^{-\Delta v_s}\right)W\left(z[\bar{t}]^*e^{-\Delta v_s}\right)}{z[t]z^*[\bar{t}]\left[1+W\left(z[t]e^{-\Delta v_s}\right)\right]\left[1+W\left(z[\bar{t}]^*e^{-\Delta v_s}\right)\right]\left[W\left(z[t]e^{-\Delta v_s}\right)-W\left(z[\bar{t}]^*e^{-\Delta v_s}\right)\right]^2}.
\end{equation}
All the information of the mixed state is encoded in this density matrix, at least in the near-horizon approximation. In particular, the diagonal of $\rho$ is the particle flux ($N_\omega$) and can be used to compute the luminosity ($L$) from
\begin{equation}
L(n)\frac{2\pi}{\epsilon}=\sum_j\hbar\omega_j N_{\omega_j}=\sum_j\hbar\omega_j\rho_{\omega_j\omega_j}.
\end{equation}
The summation over frequencies can be computed inside the integral to get
\begin{equation}
L(n)\frac{2\pi}{\epsilon}=\frac{\hbar\epsilon}{4\pi^2}\stackrel[-\infty]{+\infty}{\iint} dtd\bar{t}
G(t,\bar{t},\Delta v_s)\frac{\textrm{sinc}\left(n\pi+\left[1+W\left(z[t] \right)\right]2M\epsilon\right)\textrm{sinc}\left(n\pi+\left[1+W\left(z[\bar{t}]^*\right)\right]2M\epsilon\right)}{-2i\sin\left(\left[W\left(z[t] \right)-W\left(z[\bar{t}]^*\right)\right]2M\epsilon\right)}.\label{eq:intensidad}
\end{equation}

This integral can be solved numerically as we will show in the next section. However, due to the explicit dependence on $\Delta v_s$ (the $I^-$ distance of both shells), it is interesting to study a simplified expression corresponding to the regime where the second shell collapses long after the first one has formed the black-hole. This is explored in the next subsection.

\subsection{Late-time approximation}

Analogous to Hawking's late-time approximation, expressions (\ref{eq:numero_varz}) and (\ref{eq:intensidad}) can be simplified if one considers only late-time emissions, i.e. radiation arriving as $I^+$ close before the end of the emission due to the evaporation. Since the outgoing ray coming from $i_e$ arrives at $I^+$ at a finite retarded time $u$, instead of setting the evaporation in some distant (infinite) future, we set the evaporation time at a finite $u$ and evaluate observables pushing the early time regime to $u\to-\infty$. This is achieved by setting the second shell to collapse much later than the first one. That way any finite time $u$ before the evaporation corresponds to late-time emission. For the observables presented in the previous subsections this is implemented by taking the limit $e^{-\Delta v_s}\to0$ inside the integrals which gives 
\begin{equation}
G(t,\bar{t},0)\to\frac{1}{2\cosh\left[(t-\bar{t})/2\right]^2}\equiv\frac{1}{2}\sech\left[\frac{t-\bar{t}}{2}\right]^2 .\label{eq:G_limit}
\end{equation}
In analogy to Hawking's late-time approximation (\ref{eq:u(v)_Hawking}) it can be checked that the same result is obtained if we approximate $U(v)$ in (\ref{eq:u(v)}) by 
\begin{equation}
U_{late}(v)=v_{s2}-4M-4M\textrm{W}\left[\frac{v_{01}-v}{4M}e^{\Delta v_s}\right]\label{eq:u(v)_approx}
\end{equation}
assuming $\Delta v_s >> \frac{v_{01}-v}{4M}$. This affects the early time emission ($v<<v_{01}$) but keeps unaffected the late-time radiation ($v_{02}>v\gtrsim v_{01}$). In the next section we present a numerical comparison of the exact and approximated expressions.\\

 It is worth pointing out that the late-time expressions become independent of the relative distance of the shells. This may indicate some form of independence from the details of the evaporation. This is reminiscent of Hawking's observation that late-time emission does not depend on the details of the collapse.
 
\section{Thermal spectrum}
 
In this section we analyze the expressions for the observables computed in the previous section. We compare the full expressions with the late-time approximation by a numerical integration and we deduce the thermal part of the spectrum comparing the obtained expressions with the analogous for the Hawking radiation of a collapsing shell outlined in appendix \ref{sec:Hawking}. As a reference point we compare our results with the literature \citep{Frolov:2016gwl,Bianchi:2014bma} where the luminosity and entropy flux were computed for a nearly identical system. We begin by reviewing the technique used is those calculations.

\subsection{Conformal anomaly}\label{sec:conformal}

Any two dimensional space-time is conformally flat and the free scalar field propagating in two dimensions also exhibits Conformal invariance. This fact has been used to study Hawking radiation in two dimensional models since very early on \cite{Davies:1976ei,Christensen:1977jc}. In the four dimensional case conformal symmetry can be recovered for the geometry near the horizon \citep{Fabbri:2004yy}. The wave equation for the massless scalar field has conformal symmetry in the near-horizon approximation as well (because the centrifugal potential is ignored). Both are part of our hypothesis so we can consider this symmetry at the classical level. This is the approach taken in most of the literature to extend results obtained in two dimensional models to four dimensional space-times. It is also well known that quantization produces an anomaly in the stress-energy tensor in the sense that, under a conformal transformations its expectation value changes not only in a conformal (multiplicative) factor but also includes a non zero additive term. Since the \textit{in} and \textit{out} constructions presented in section I can be connected by a conformal transformation, the knowledge of the expectation value in one of them (say $\left\langle T_{ab}\right\rangle_{in}=0$) allows to compute the other ($\left\langle T_{ab}\right\rangle_{out}$) through the calculation of the "anomalous" contribution. This has been done in the literature \citep{Davies:1976ei,Christensen:1977jc,Fabbri:2004yy,Bianchi:2014bma} for several models, including this \textit{sandwich} model of an EBH \citep{Hiscock:1980ze,Frolov:2016gwl}.

\begin{figure}
\centering
\subfloat[]{\includegraphics[scale=0.6]{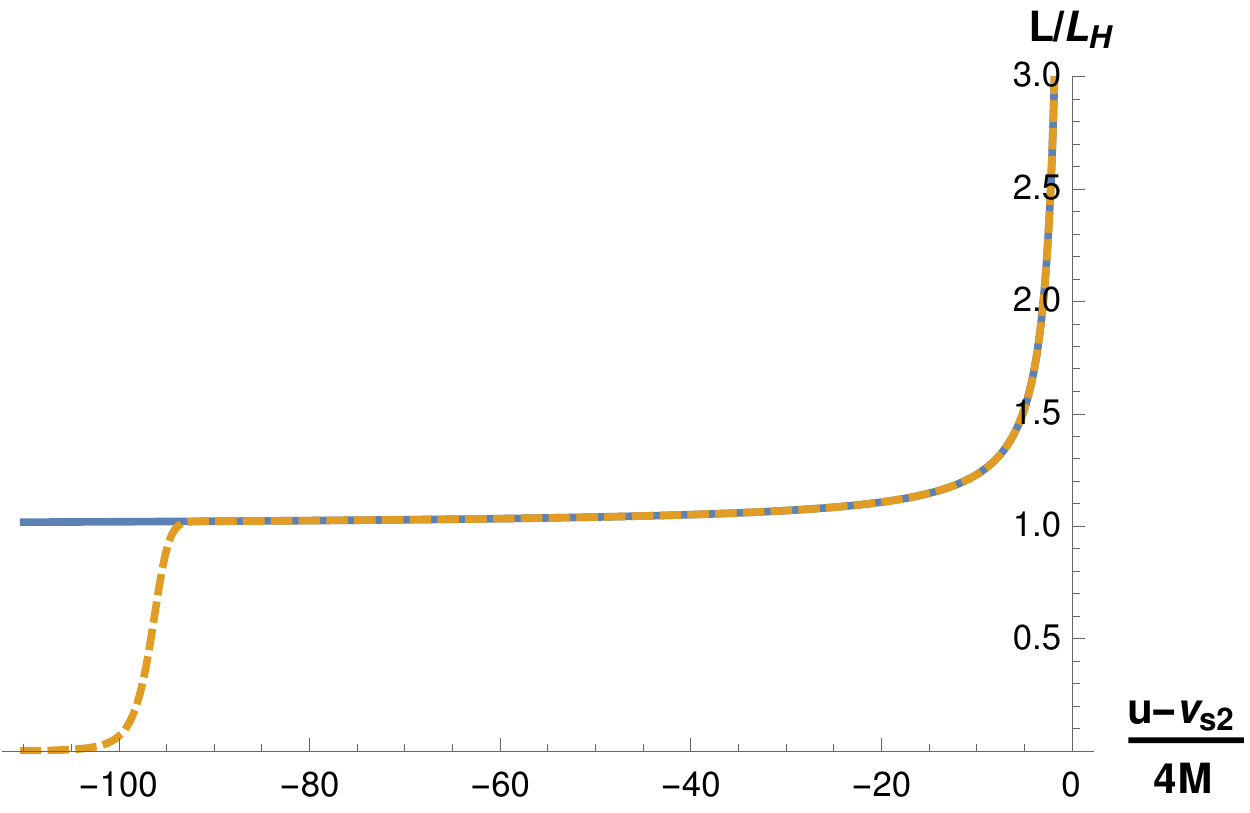}}
\qquad
\subfloat[]{\includegraphics[scale=0.6]{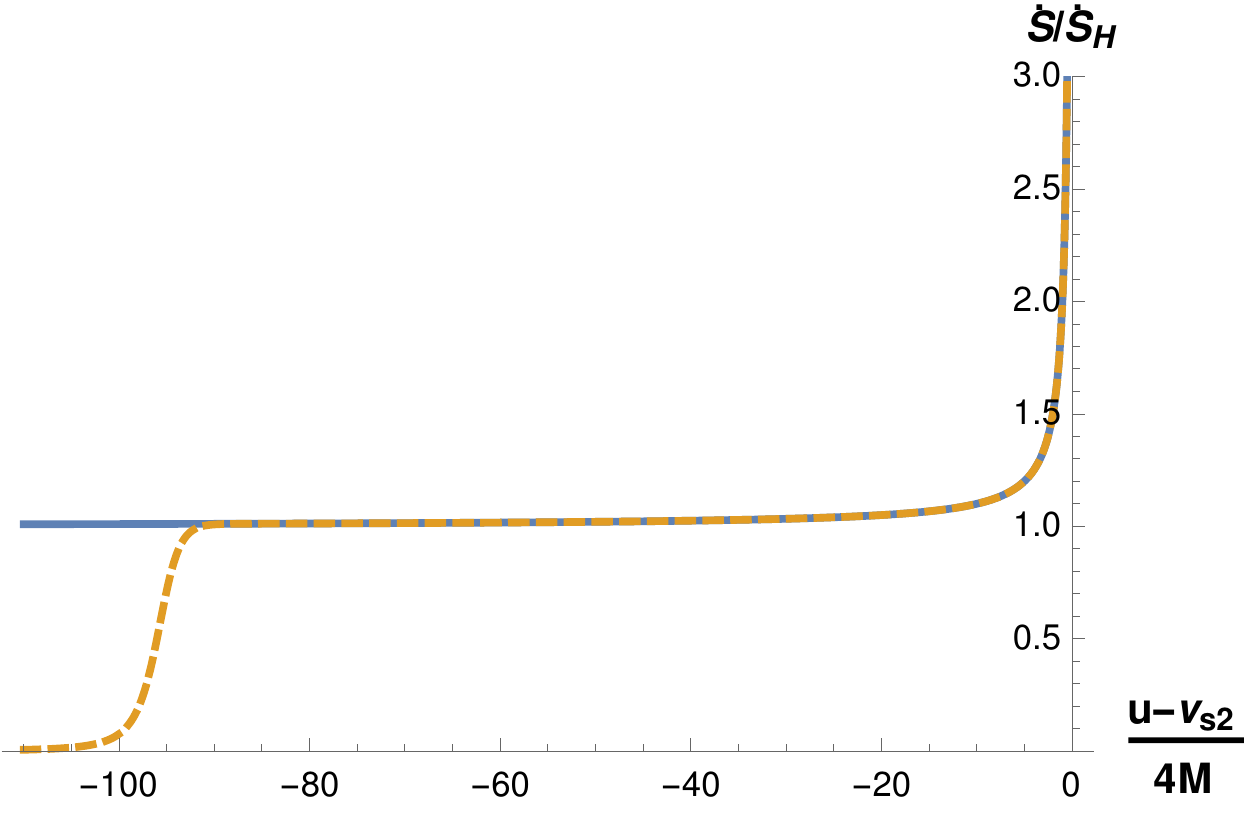}}\caption{Conformal anomaly calculation of the luminosity (a) and entropy flux (b) in units of late-time Hawking flux ($L_H$ and $\dot{S}_H$ respectively). Dashed lines show the beginning of emission corresponding to an interval $4M\Delta v_s$ before evaporation and the solid line is the late-time approximation ($\Delta v_s\to+\infty$). Both luminosity and entropy flux diverge at the evaporation time. Plot is done for $\Delta v_s=100$.}\label{fig:anomalos}
\end{figure}

Following the notation of section II.D in \citep{Frolov:2016gwl} we start from the quantity

\begin{equation}
P=\ln\left|\frac{dV(u)}{du}\right|
\end{equation}
where $V(u)$ is the inverse of $U(v)$ given by (\ref{eq:u(v)}). Two of the observables that can be computed from this quantity are the luminosity \begin{equation}
L(u)=\frac{\hbar}{48\pi}\left[-2\frac{d^2 P}{du^2}+\left(\frac{dP}{du}\right)^2\right]\label{eq:conformal_energy}
\end{equation}
and the entropy flux
\begin{equation}
\dot{S}(u)=-\frac{\hbar}{12}\dot{P}.
\end{equation}
Figure \ref{fig:anomalos} shows both observables as functions of $u$ for the exact expression (\ref{eq:u(v)}) and also for the late-time approximation (\ref{eq:u(v)_approx}). In the latter, the explicit expression are very simple
\begin{eqnarray}
P_{late}(x)&=&\ln|1+x|+x-\Delta v_s\\
\dot{S}_{late}(x)&=&\frac{\hbar}{48 M}\frac{x+2}{x+1}=\dot{S}_H\frac{x+2}{x+1}\label{eq:Sconformal}\\
L_{late}(x)&=&\frac{\hbar}{768 M^2}\frac{(x+2)^2+2}{(x+1)^2}=L_H\frac{(x+2)^2+2}{(x+1)^2}.\label{eq:Iconformal}
\end{eqnarray}
where $x=\frac{v_{s2}-4M-u}{4M}>-1$. All this expressions present a coincidence with Hawking radiation at the far past ($x\to+\infty$) and a divergence at the evaporation time ($x\to -1$).

\subsection{Luminosity computation}
\begin{figure}
\centering
\subfloat[{Sandwich BH with $\Delta v_s=100$. There is emission during an interval $4M\Delta v_s$.}]{\includegraphics[scale=0.4]{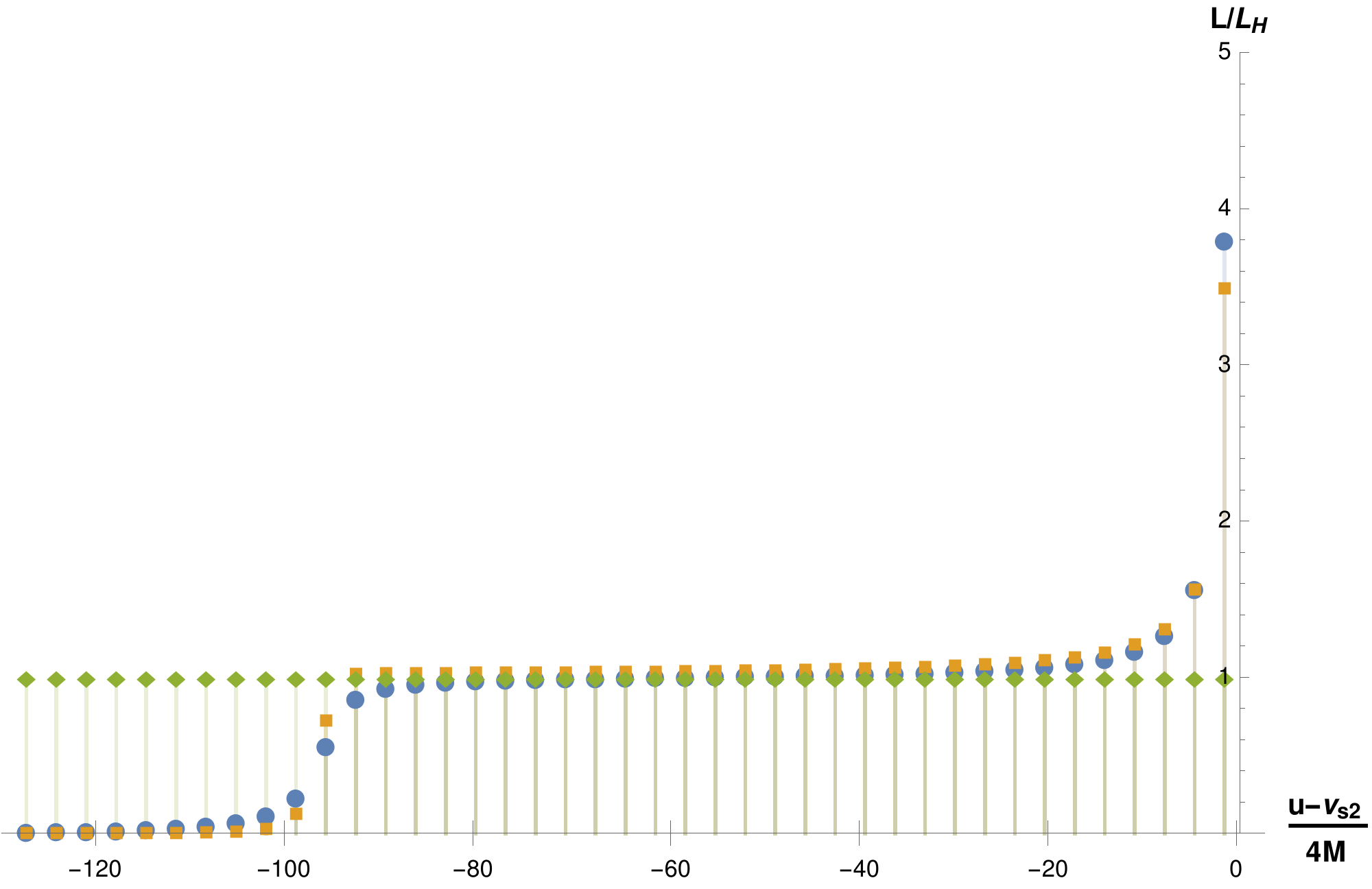}\label{fig:transitorio}}
\qquad
\subfloat[{Late-time approximation ($\Delta v_s\to\infty$). Radiation slowly departs from thermal profile. }]{\includegraphics[scale=0.4]{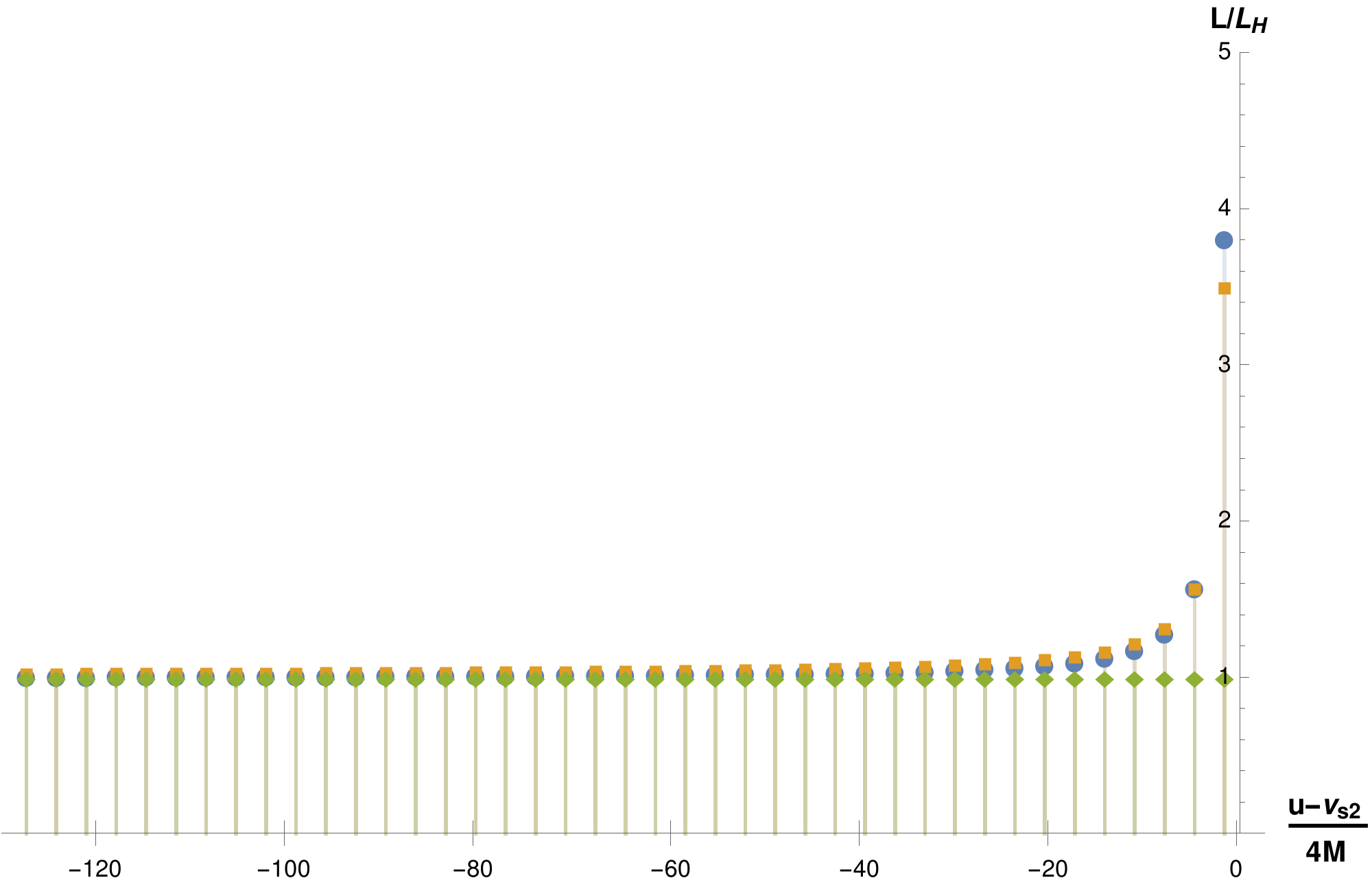}\label{fig:anomalia_aprox}}\caption{\small Luminosity according to the conformal anomaly calculation (orange squares) and numerical computation of (\ref{eq:intensidad}) (blue circles). Both measured with respect to late-time Hawking luminosity ($L_H$, green rhombi). There is qualitative agreement between conformal anomaly and Bogoliubov transformation technique but quantitative discrepancies appear because the later method require wave packets extended in frequency and time. Agreement is better in (b) for the same reason.}
\end{figure}

Numerical computation of the luminosity from expression (\ref{eq:intensidad}) shows a very good agreement with  (\ref{eq:conformal_energy}) as is shown in Figure [\ref{fig:transitorio}]. The qualitative features of the two curves coincide but there are two effects causing quantitative differences. One is the truncation of the integration range to achieve convergence of the numerical integral. The other, more fundamental, is the use of non-local wave packets to compute the Bogoliubov coefficients which causes the curve to be smoother. This is an inescapable feature of our method.

For the late-time approximation, shown in Figure \ref{fig:anomalia_aprox}, agreement between the numerical integrals and the conformal anomaly computation improves because it lacks a sharp step to register in the tail of the wave packets. Disagreement appears towards the evaporation time where the non-local nature of the wave packets is most apparent. An important feature of this approximation is the fact that there is a slow deviation from Hawking radiation which plays the role of asymptotic limit to the past. This is well captured by both computations of the luminosity.


\subsection{Density matrix and thermality}\label{sec:matriz_densidad}

\begin{figure}
\centering
\subfloat[Thermal and total contribution to the total luminosity with respect to usual Hawking radiation. Their ratio remains close to 1 for all times.]{\includegraphics[scale=0.3]{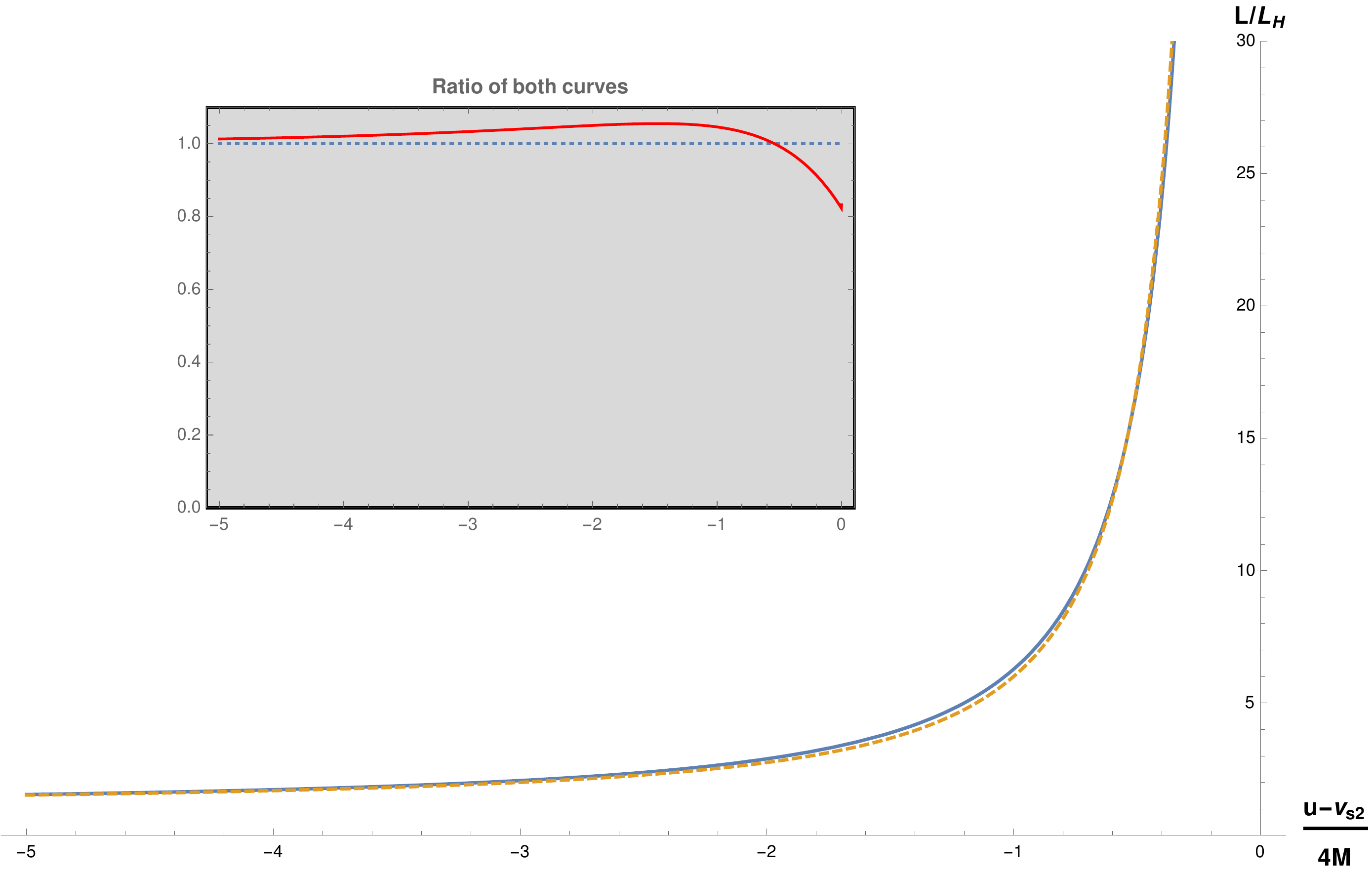}}
\qquad
\subfloat[Thermal and total contribution to the total entropy flux with respect to Hawking flux $\dot{S}_H$. The ratio stay close to 1 but approaches $\pi/2$ towards evaporation.]{\includegraphics[scale=0.3]{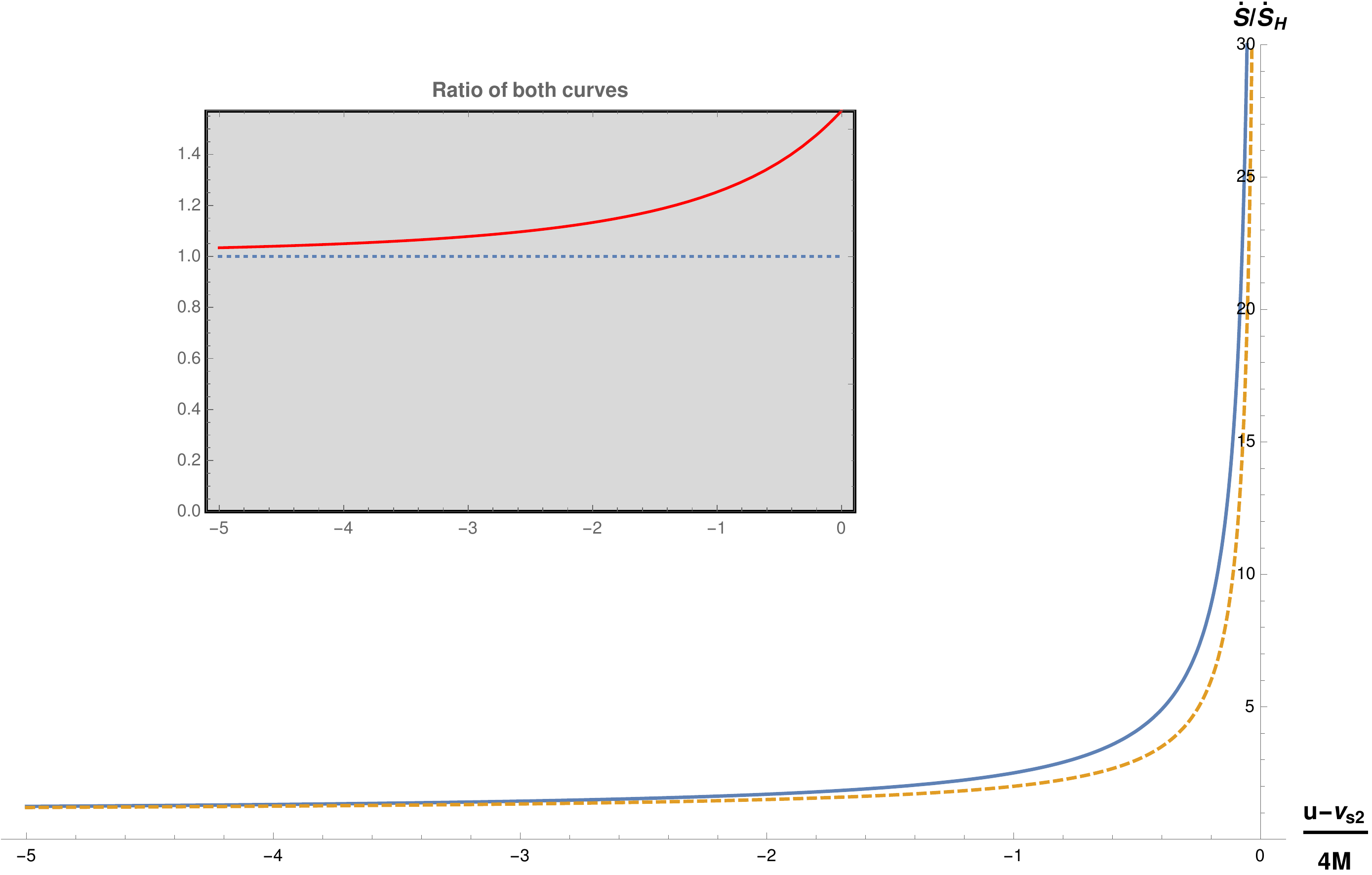}}\caption{Thermal contribution to the radiation with respect to Hawking's late-time radiation. Blue solid line shows the thermal contribution (according to our calculation), orange dashed is the total contribution according to conformal anomaly and solid red in the smaller plot is their ratio over time.}\label{fig:thermalcontribution}
\end{figure}

Numerical evaluation of expression (\ref{eq:numero_varz}) for the density matrix is possible but the use of wave-packet obscures the result. However it is useful to compare the integrand in the late-time limit (\ref{eq:G_limit}), where the density matrix reduces to
\begin{eqnarray}
\rho_{\omega_j\omega_k}(n)&=&\frac{1}{(4\pi)^2}\frac{\epsilon}{\sqrt{\omega_j\omega_k}}\stackrel[-\infty]{+\infty}{\iint} 
dtd\bar{t}A(u_n,t,\bar{t})B(\omega_j,\omega_k,t,\bar{t}),\\
A(u_n,t,\bar{t})&=&\textrm{sinc}\left(\left\lbrace\frac{u_n-v_{s2}}{4M}+\left[1+W\left(z[t] \right)\right]\right\rbrace 2M\epsilon\right)\textrm{sinc}\left(\left\lbrace \frac{u_n-v_{s2}}{4M}+\left[1+W\left(z[\bar{t}]^*\right)\right]\right\rbrace 2M\epsilon\right)\sech\left[\frac{t-\bar{t}}{2}\right]^2,\\
B(\omega_j,\omega_k,t,\bar{t})&=&\exp\left(i4M\omega_j\left[1+W\left(z[t] \right)\right]-i4M\omega_k\left[1+W\left(z[\bar{t}]^*\right)\right]\right),
\end{eqnarray}
and the analogous integral (\ref{eq:matriz_densidad_H}) for late-time Hawking radiation
\begin{eqnarray}
\rho^H_{\omega_j\omega_k}&=&\frac{1}{(4\pi)^2}\frac{\epsilon}{\sqrt{\omega_j\omega_k}}\stackrel[-\infty]{+\infty}{\iint} 
dtd\bar{t}A^H(u_n,t,\bar{t})B^H(\omega_j,\omega_k,t,\bar{t}),\\
A^H(u_n,t,\bar{t})&=&\textrm{sinc}\left(\left[\frac{u_n-v_{0}}{4M}+t+\frac{\pi}{2}i\right]2M\epsilon\right)\textrm{sinc}\left(\left[\frac{u_n-v_{0}}{4M}+\bar{t}-\frac{\pi}{2}i\right]2M\epsilon\right)\sech\left[\frac{t-\bar{t}}{2}\right]^2,\\
B^H(\omega_j,\omega_k,t,\bar{t})&=&\exp\left(-2M[\omega_j+\omega_k]\pi+i4M[\omega_j t-\omega_k\bar{t})\right].
\end{eqnarray}
Inspecting both integral expressions is clear that $A$ and $B$ coincide with $A^H$ and $B^H$ under the replacement
\begin{eqnarray}
1+Re\left[W\left(z[t]\right)\right]&\to& t,\\
1+Re\left[W\left(z[\bar{t}]^*\right)\right]&\to& \bar{t},\\
Im\left[W\left(z[t]\right)\right]&\to&\frac{\pi}{2},\\
Im\left[W\left(z[\bar{t}]\right)^*\right]&\to&-\frac{\pi}{2},\\
u_n-v_{s2}&\to& u_n-v_0.
\end{eqnarray}
\begin{figure}
\centering
\subfloat[For early time (solid orange) the main contribution comes from the thermal spectrum ($t=\bar{t}>>1$). For late times (dot-dashed red) a non thermal contribution from evaporation ($t=\bar{t}<0$) takes over the integral. Thermal contribution grows with time and the visual indicator is the growing peak from the asymptotic value $\exp(4\pi M\omega_j)$.]{\includegraphics[scale=0.4]{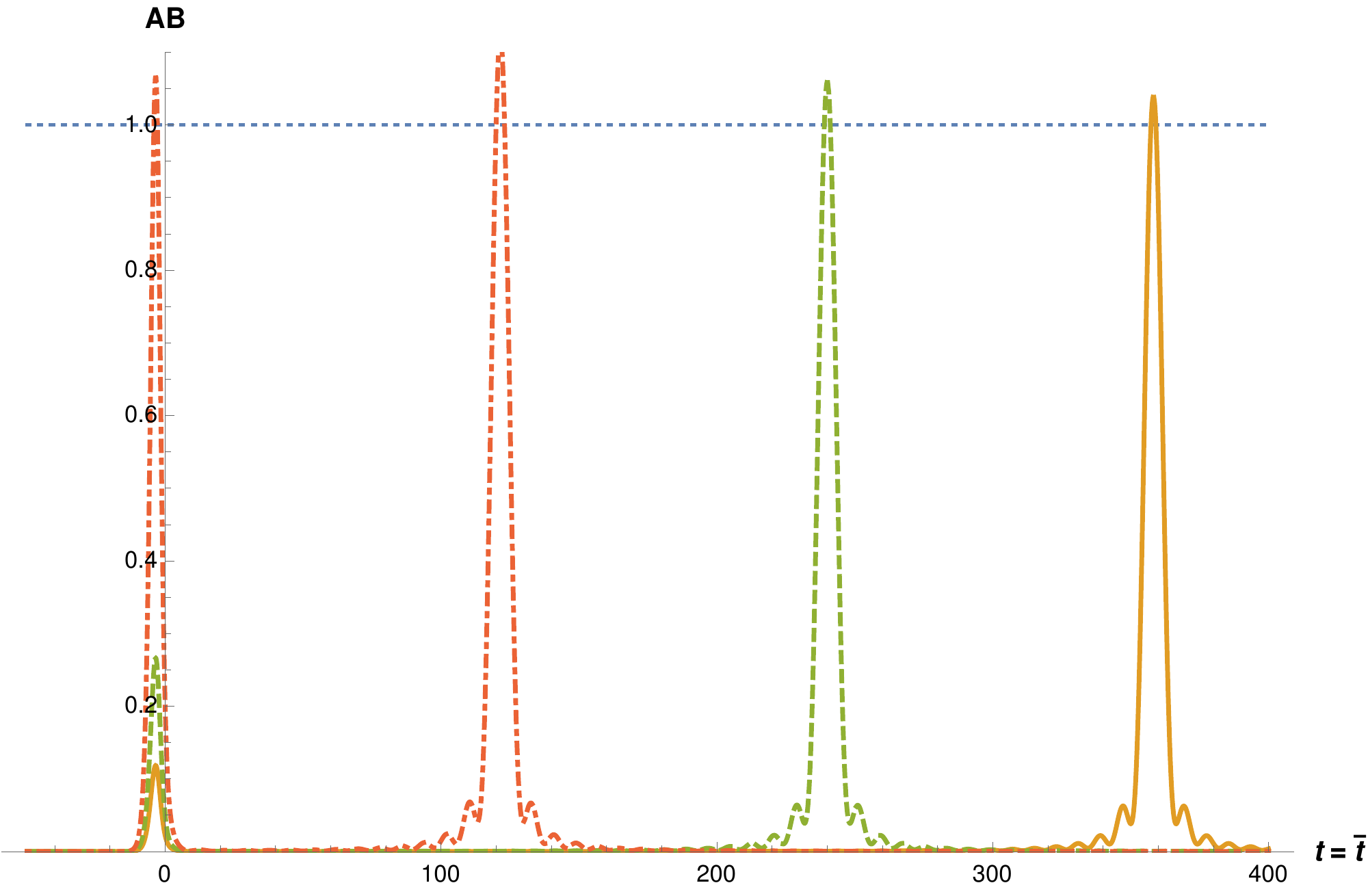}}
\qquad
\subfloat[Integrand for Hawking's thermal spectrum presented for comparison. The integrand is time independent ($t$ and $\bar{t}$ can be shifted the same amount without changes to the integral). In particular its peak takes the value $\exp(4\pi M\omega_j)$ fixing the temperature.]{\includegraphics[scale=0.4]{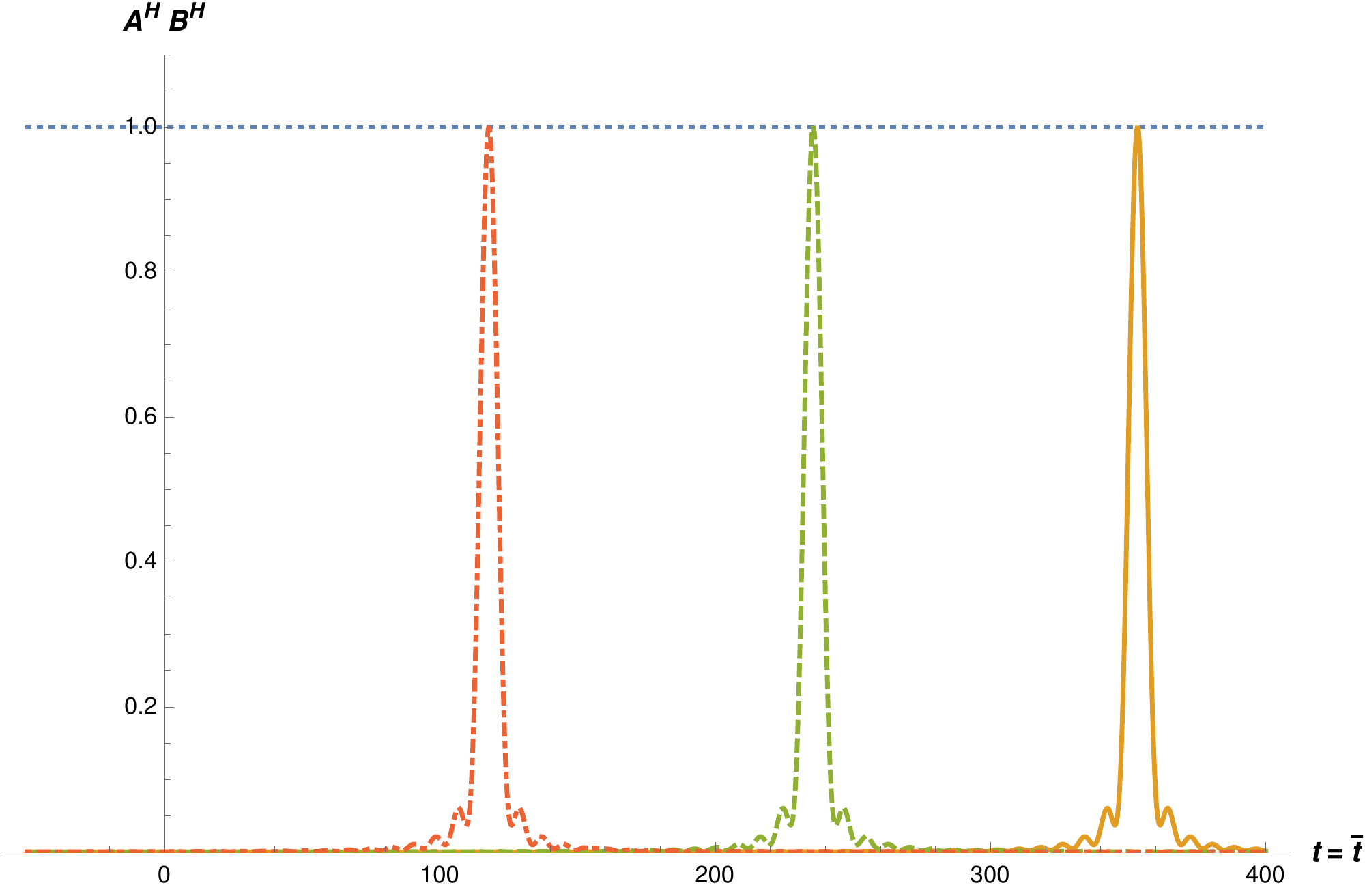}}\caption{Integrand in the calculation of $N_{\omega_j}(u_n)$ (the particle flux for frequency $\omega_j$ at advanced time $u_n$) evaluated over the line $t=\bar{t}$ in the case $M\epsilon=0.2$, $j=5$ and three different times: $n=-45$ (solid orange), $n=-30$ (dashed green) and $n=-15$ (dot-dashed red). Everything is normalized by the maximum of $A^H B^H$ given by $\exp(4\pi M\omega_j)$.}\label{fig:integrandos}
\end{figure}
This is the asymptotic limit when $t,\bar{t}\to+\infty$ as is shown in Figure \ref{fig:W}. Therefore, the contributions from that sector of both integrals match. This can be seen by comparing Figures \ref{fig:integrandos}.a and \ref{fig:integrandos}.b where the special case $\omega_j=\omega_k$ is plotted.\footnote{Since $\rho^H$ is independent of $u_n$ the explicit parameters $\frac{u_n-v_{s2}}{4M}$ and $\frac{u_n-v_0}{4M}$ in both integral expressions can be chosen to coincide without lose of generality.}. On the other hand, the integrand is quite different in the limit $t,\bar{t}\to-\infty$ and this sector needs to be carefully considered. Comparing the functions $A$ and $A^H$ we see their amplitudes are bump functions with exponential decay in the $t+\bar{t}=const.$ directions and exhibit a $\textrm{sinc}$ function decay in the $t-\bar{t}=const.$ directions. For both functions the main contribution to the integrals comes from the $t=\bar{t}$ line with a peak when the conditions 
\begin{equation}
\frac{u_n-v_{s2}}{4M}+1+Re\left[W\left(z[t]\right)\right]=0,\label{eq:Hawkingfactor_time}
\end{equation}
for $A$ and
\begin{equation}
\frac{u_n-v_{s0}}{4M}+t=0\label{eq:Hawkingfactor_timeH}
\end{equation}
for $A^H$ are meet. As mentioned in the appendix, the density matrix (\ref{eq:matriz_densidad_H}) is time independent, so the condition (\ref{eq:Hawkingfactor_timeH}) has no physical significance. This is not the case for condition (\ref{eq:Hawkingfactor_time}) which encodes the time dependency of the density matrix $\rho_{\omega_j\omega_k}(n)$.
Comparison of $B$ and $B^H$ has a more nuance interpretation. $B^H$ has a constant amplitude
\begin{equation}
|B^H(\omega_j,\omega_k,t,\bar{t})|=\exp\left(-2M\pi[\omega_j+\omega_k]\right)
\end{equation}
but $B$ has an amplitude representing a two dimensional step function between two asymptotic values
\begin{equation}
|B(\omega_j,\omega_k,t,\bar{t})|=\exp\left(-4M\omega_jIm[W(z[t])]-4M\omega_k Im[W(z[\bar{t}]^*)]\right)=\left\{ \begin{array}{lcl}
|B^H(\omega_j,\omega_k)|,\qquad t,\bar{t}\to +\infty\\
\qquad\qquad\quad 1,\qquad t,\bar{t}\to -\infty\\
\end{array}\right.
\end{equation}
This step can give rise to a second relevant contribution to the integral depending on the location of the peak in $A$ since the density matrix is the integral of the product $AB$ (see again the comparison of Figures \ref{fig:integrandos}.a and \ref{fig:integrandos}.b).
For early times ($u_n-v_{s2}<<0$) the peak in $A$ is located in the $t=\bar{t}>>1$ sector and the second contribution is suppressed. Here is where the thermal spectrum is obtained. For late times ($u_n-v_{s2}\sim 0$) the step in $B$ coincides with the peak in $A$ and this second contribution dominates. Even when the contribution from the step function in $|B|$ is not dominant, the result of the integral is time dependent because the Hawking factor $e^{-2M\pi(\omega_j+\omega_k)}$ is replaced by
\begin{equation}
\exp\left(-4M\omega_jIm[W(z[t])]-4M\omega_kIm[W(z[\bar{t}]^*)]\right),
\end{equation}
which reduces to 
\begin{equation}
\exp\left(-8M\omega_jIm[W(z[t])]\right)
\end{equation}
over the curve $t=\bar{t}$. Evaluating this factor at the peak given by (\ref{eq:Hawkingfactor_time}) we can define a time dependent factor, independent of the integration variables $t,\bar{t}$ that provide us with a first approximation for the density matrix in the evaporation scenario. In particular, such approximation neglects the second contribution to the integral mentioned before. The computation requires to write the imaginary part ($IW$) of the function $W\left(z[t]\right)$ in terms of its real part ($RW$). By definition of $z[t]$ in (\ref{eq:z_t}) and the defining property of the Lambert $W$ function we have
\begin{equation}
W(z[t])e^{W(z[t])}=z[t]=-e^{-1}+ie^t.
\end{equation}
The real part of this equation 
\begin{equation}
e^{RW}\left(RW\cos\left[IW\right]-IW\sin\left[IW\right]\right)=-e^{-1}\label{eq:implicit}
\end{equation}
is an implicit solution for $IW(RW)$ and then (\ref{eq:Hawkingfactor_time}) provide an implicit expression for $IW(u_n)$. The density matrix obtained by the substitution
\begin{equation}
4\pi M\omega\to 8 M\omega IW(u_n),
\end{equation}
in Hawking's expression for the density matrix of thermal radiation is a good approximation for early times in the evaporation scenario. This can be interpreted as the substitution of Hawking temperature ($T_H$) by an effective, time dependent, temperature
\begin{equation}
T(u)=T_H\frac{\pi/2}{IW(u)}.
\end{equation}
For this effective temperature, the luminosity of a thermal profile would be
\begin{equation}
I(u)=I_H\left(\frac{\pi/2}{IW(u)}\right)^2
\end{equation}
and the entropy flux
\begin{equation}
\dot{S}(u)=\dot{S}_H\frac{\pi/2}{IW(u)}.
\end{equation}
Numerical comparison of this expressions with (\ref{eq:Sconformal}) and (\ref{eq:Iconformal}) present an excellent agreement for the luminosity profile and a slightly worse agreement for the entropy as is shown in Figure \ref{fig:thermalcontribution}. This indicates that the divergence in both quantities is driven by a growing temperature thermal spectrum and most of the energy is emitted as thermal spectrum. However entropy is more sensitive to the non-thermal part of the spectrum. A curious result is that the ratio of both calculation of entropy flux tends to $\pi/2$ towards evaporation. This hints to interesting physics to be explored, relating the thermal and non thermal part of the spectrum towards evaporation.

\section{Related models with well defined Cauchy problem}

The quantization of the scalar field presented in section \ref{sec:emision} requires extra assumptions in order for the initial value problem to be well posed. In particular, the extension to region \textit{V} (after the evaporation) is not automatic because it is beyond the Cauchy horizon $C$. However, there are models with a well defined Cauchy problem in the region of interest closely related to the $EBH$. If the initial $BH$ loses most of its mass but a small $BH$ remains, then the singularity stays within the (finite size) event horizon and no evaporation point ($i_e$) nor Cauchy horizon ($C$) forms. The $EBH$ can be obtained as a zero mass limit of such remnant $BH$ ($RBH$) as we will discuss in the first subsection. A well posed Cauchy problem can also be obtained in the so called \textit{regular Black Hole} ($RegBH$) models, where the region of high curvature of the $EBH$ is replaced by a (Planck size) regular metric. Here, the Cauchy problem for a matter field is well posed in the entire space-time because there isn't a singularity or event horizon, which is replaced by an inner apparent horizon.

\subsection{Remnant Black Hole}

\begin{figure}
\centering
\subfloat[]{\includegraphics[height=8cm]{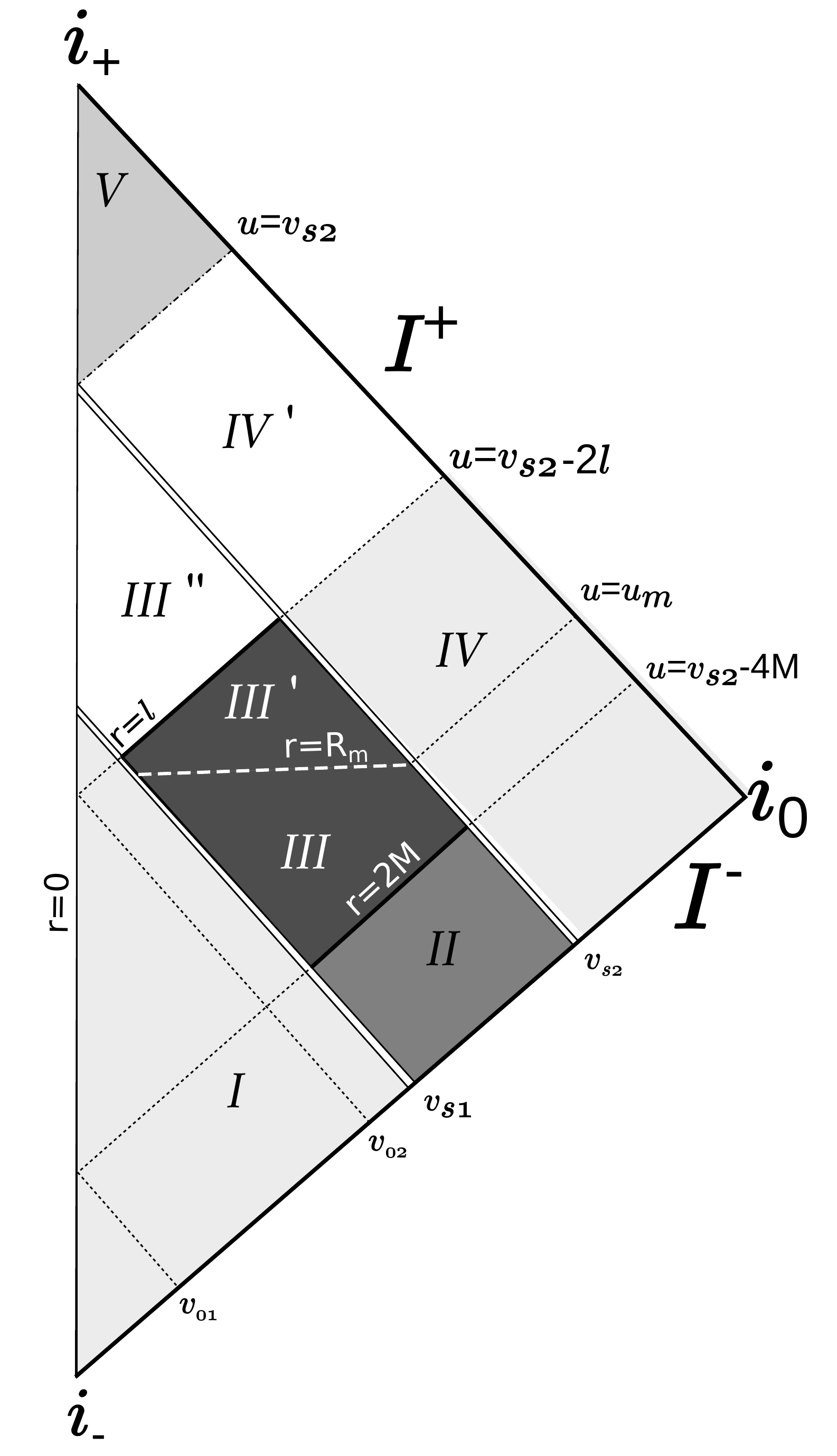}\label{fig:espacio_tiempo_deSitter}}
\qquad
\subfloat[]{\includegraphics[height=8cm]{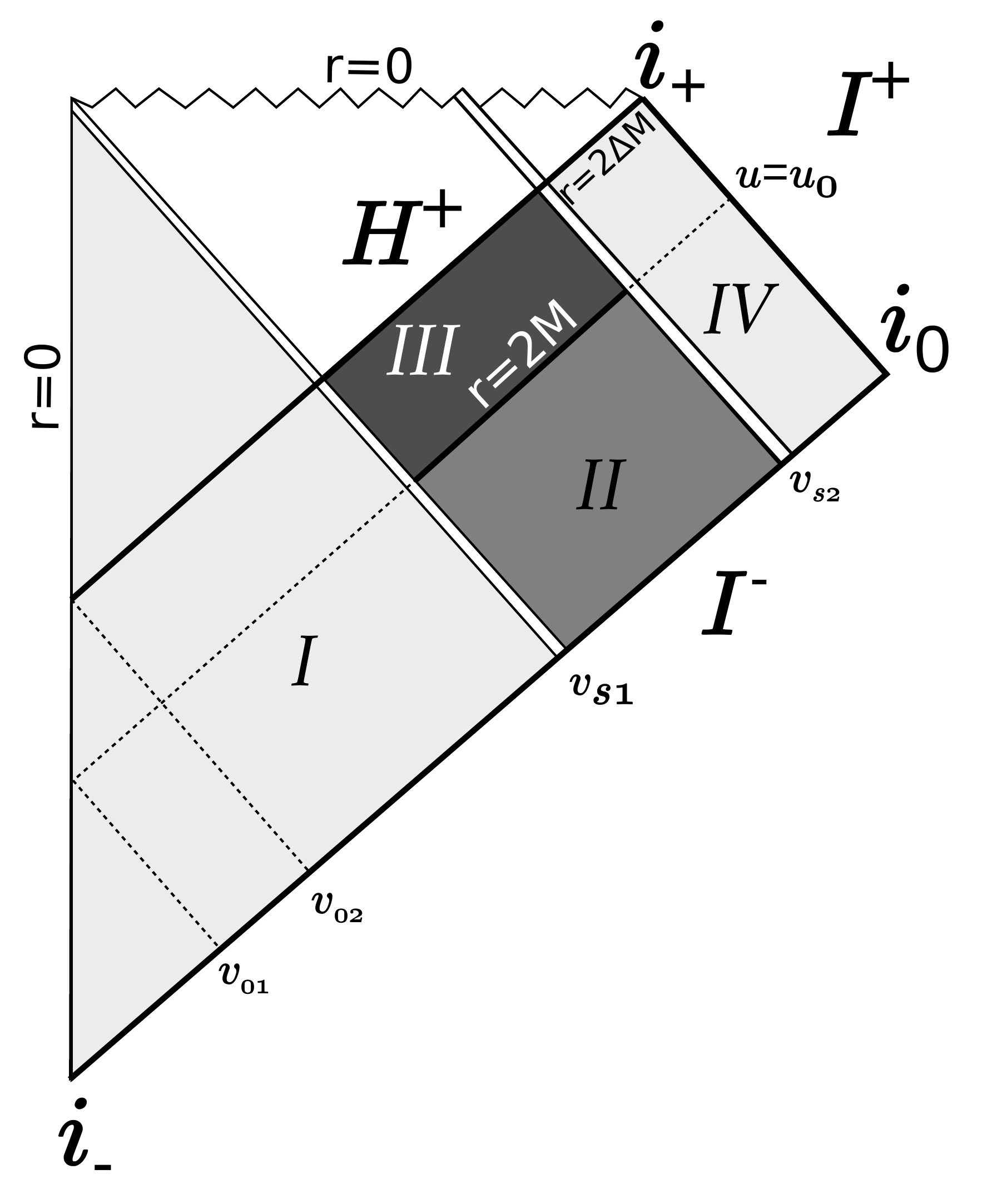}\label{fig:espacio_tiempo_remanente}}\caption{Penrose-Carter diagrams of two models with well defined Cauchy problem for the matter fields. The $EBH$ model (\ref{eq:metrica_remanente}) can be seen as a limit of each of this models.\\
(a) Regular Black Hole (\ref{eq:metrica_regular}): regions \textit{I,II,IV} and \textit{V} are the same as for the $EBH$ (Figure \ref{fig:sandwich}). Region \textit{III} is replaced by \textit{III}' for $l<r<R_m$. Region \textit{III}'' replaces the interior of the horizon $H^+$ and the Cauchy horizon $C$ is replaced by the interface region \textit{IV}'.The limit $R_m,l\to 0$ leads to the $EBH$ case.\\(b)Remnant metric (\ref{eq:metrica_remanente}) : regions $I$ to $III$ are the same as for the $EBH$. Region $IV$ is the exterior of a remnant BH of mass $\Delta M$ and the region beyond evaporation is not present. The limit $\Delta M\to 0$ leads to the $EBH$ scenario.}
\end{figure}

Lets consider a remnant Vaidya metric ($RBH$)
\begin{equation}
ds^{2}=-\left(1-\frac{2M(v)}{r}\right)dv^{2}+2dvdr+r^{2}d\Omega^{2},\label{eq:metrica_remanente}
\end{equation}
where\begin{equation}
M(v)=M\theta(v-v_{s1})-(M-\Delta M)\theta(v-v_{s2}).
\end{equation}
The space-time is identical to the EBH discussed in previous sections for regions $I,II \& III$. Region \textit{IV} is replaced by the metric of a remnant BH with mass $\Delta M$ and region \textit{V} is missing (see Figure \ref{fig:espacio_tiempo_remanente}).

\begin{figure}
\includegraphics[height=7cm]{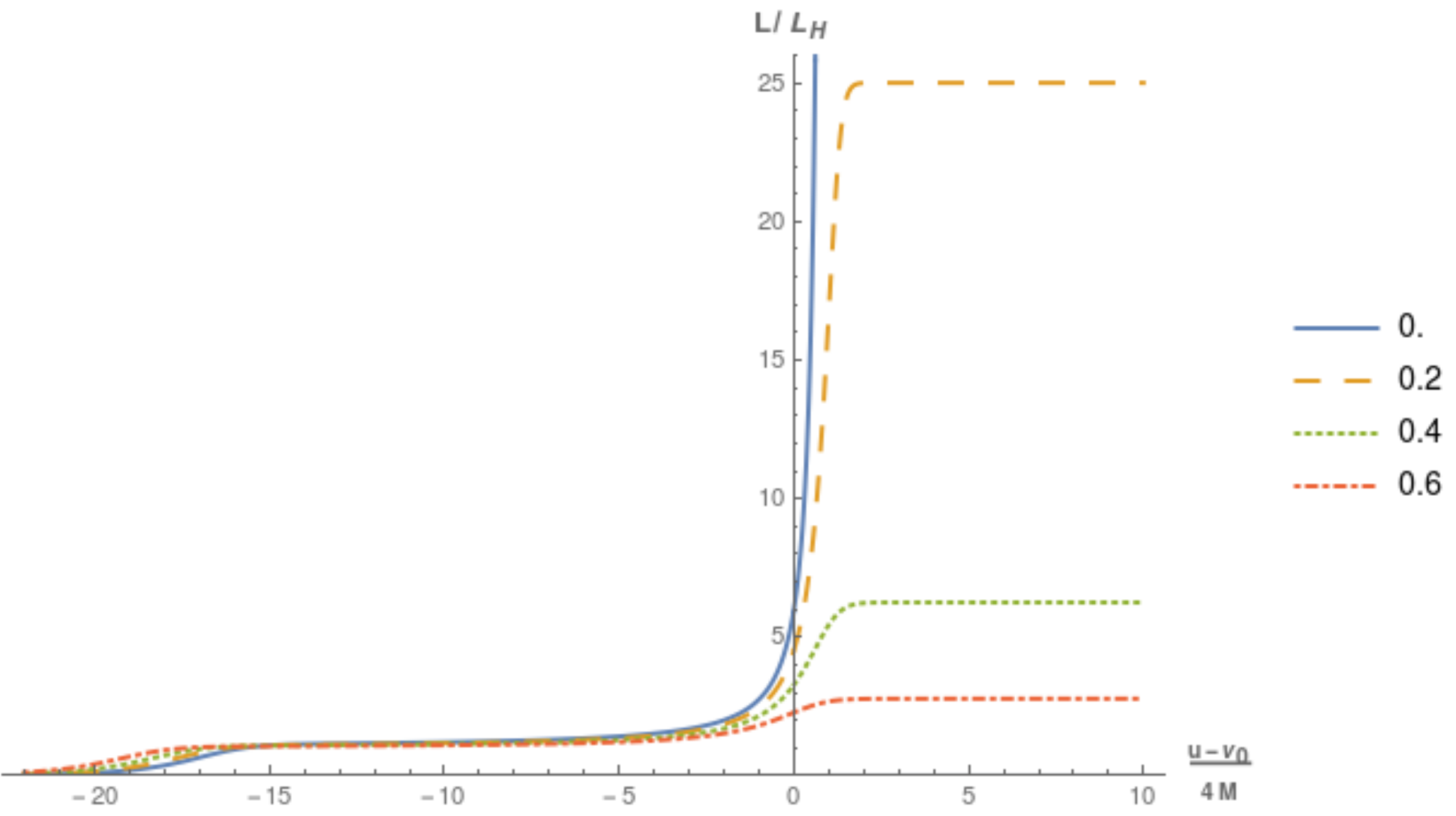}
\caption{Dimensionless luminosity emitted according to the conformal anomaly calculation. Solid blue line is for total evaporation at $u=v_0+4M$ where the luminosity diverges. Dashed and dotted lines are remnant scenarios with different values for the parameter $\Delta M/M=0.2,0.4,0.6$. They present a final late-time thermal radiation interpolating between the EBH ($\Delta M/M=0$) and no evaporation ($\Delta M/M=1$) scenarios. Plot is done for $\Delta v_s =20$ (in units of $4M$) to see the start of the radiation in the plot.}\label{fig:energia_remanente}
\end{figure}

This geometry provides a well posed Cauchy problem for the scalar field outside the Horizon ($H^+$) and the Bogoliubov coefficients for plane waves are analogous (under the near-horizon approximation) to the ones computed from (\ref{eq:beta}) for modes emitted prior to evaporation. The only change is the substitution of the $U(v)$ function by
\begin{equation}
U_{remnant}(v)= U(v)-4\Delta M\ln\left[\frac{v_{s2}-U(v)-4\Delta M}{4M_0}\right],\label{eq:u_rem}
\end{equation}
representing the arrival time at $I^+$ of a ray that leaves $I^-$ at $v$. The function $U(v)$ is the $\Delta M\to 0$ limit in expression (\ref{eq:u_rem}). On the other end, the scenario without evaporation discussed in appendix A is obtained at the $\Delta M\to M$ limit where $U_{remnant}(v)\to U_H(v)$. The Bogoliubov coefficients and the density matrix are also obtained as limits of the $RBH$ but unfortunately the computation of the Bogoliubov coefficients in the $RBH$ scenario is much more convoluted than in the limits. However, we can use conformal anomaly techniques to have a qualitative understanding of the transition between both models when $\Delta M\to 0$. For example, the luminosity for the $RBH$ can be computed through (\ref{eq:conformal_energy}) and compared to the $EBH$ model. The result, presented in Figure \ref{fig:energia_remanente} as a succession of plots for decreasing $\Delta M/M$, shows that the profile described earlier has a natural interpretation in the $RBH$. It represents the intermediate stage connecting the early Hawking radiation of a $BH$ with mass $M$ and the final (late-time) radiation of a remnant with mass $\Delta M\to 0$. In this picture the increasing effective temperature found in the $EBH$ is an interpolation between the initial an final Hawking temperatures, in the special case where the final temperature diverges. The evaporation time is interpreted as the time of thermalization to the remnant's Hawking radiation. Notice that the near-horizon approximation in region $IV$ is valid far from the remnant horizon, i.e, before evaporation time. Outgoing wave-packets traveling near the event horizon face a centrifugal potential with a maximum of order $\Delta M^{-2}$ located roughly at  $r=3\Delta M$. In the limit $\Delta M\to 0$ the maximum diverges and the high potential region is pushed to $r=0$. Unlike the $EBH$ model, there is a second region of late-time Hawking radiation. The near-horizon approximation can be used to compute the thermal spectrum, although here it means to neglect the high potential near the event horizon of the remnant. This region is replaced by region $V$ in the $EBH$ case where no radiation is generated.

\subsection{Regular Black Hole}
Motivated by Quantum Gravity proposals (see a discussion of models in \citep{Frolov:2016pav}), there has been recent interest in space-times that resemble $BH$ solutions up to some arbitrary (Planck) scale and the singularity is replaced by a \textit{quantum region} of finite curvature. An example discussed in \citep{Bianchi:2014bma} is the metric
\begin{equation}
ds^2=-F(r,v)dv^2+2dvdr+r^2d\Omega^2\label{eq:metrica_regular}
\end{equation} 
where
\begin{equation}
F(r,v)=\left\lbrace\begin{array}{lcl}
1,\qquad\qquad\qquad v<v_{s1}\\
1-2M/r,\qquad \left\lbrace v_{s2}>v>v_{s1},r>R_m\right\rbrace\\
1-(r/l)^2,\qquad \left\lbrace v_{s2}>v>v_{s1},r\leq R_m\right\rbrace\\
1,\qquad\qquad\qquad v>v_{s2}.
\end{array}\right.
\end{equation} 
As is shown in Figure \ref{fig:espacio_tiempo_deSitter} regions \textit{I,II} and \textit{IV} are the same as in the $EBH$ and there is an interface (\textit{IV'}) between regions \textit{IV} and \textit{V}. Region \textit{III} is the same for $r>R_m$ and is replaced by a de Sitter patch (\textit{III'} and \textit{III''}) with cosmological constant $l^{-2}$ for $r<R_m$. The event horizon is replaced by an inner apparent horizon at $r=l$ that serves as limit of the trapped surfaces region.

Under the near-horizon approximation only the radiation arriving to $I^+$ at $u>u_m$ knows about the quantum region and this is beyond $u=v_{s2}-4M$. Therefore, considering wave packets with $M\epsilon\sim 0.1$ as we have done for Figures \ref{fig:transitorio} and \ref{fig:anomalia_aprox}, the difference between this model and the $EBH$ would only be apparent for the last wave packet ($n=-1$) before evaporation. The main feature of the radiation computed from this and similar models (see \citep{Bianchi:2014bma,Frolov:2016gwl}) is the lack of divergence in observables like the luminosity and the entropy flux. They start growing according to the $EBH$ model and have a sharp decline close before the evaporation time. In model (\ref{eq:metrica_regular}) this is controlled by the Planck scale parameters $l$ and $R_m$ and the limit $R_m,l\to 0$ is the way to recover the $EBH$ results.

\section{Conclusions}

We have successfully applied the Bogoliubov transformation approach to compute Hawking radiation profile in the case of a toy model EBH. Previous results obtained by other techniques coincide in the near-horizon approximation. However with this method we are able to identify the thermal part of the spectrum and compute an effective temperature that grows with time from the Hawking temperature of the BH formed by the collapse. Comparing the result with the limit of a remnant scenario where the mass of the final BH goes to zero we find a clear interpretation for the effective temperature. It interpolates between the initial an final temperatures but the final one is not finite in the $EBH$ model. The divergence of several observables at evaporation can be traced to this infinite final temperature.

Even when we are using a simple toy model there is qualitative agreement with the semi-classical picture where the quasi-stationary evaporation of the BH produces radiation at an increasing temperature. This indicates that some features of late-time evaporation are quite general. The fact that the $EBH$ can be obtained as the limit of other toy models with similar radiation profiles suggests this is indeed the case.
 
Besides the thermal spectrum, the Bogoliubov coefficients method encodes the non-thermal part of the spectrum. We have not characterized that contribution yet, but the limitations are mostly technical. This non-thermal contribution grows from the beginning of the radiation process and do not appear in the quasi-stationary description of the radiation process. It would be interesting to explore whether or not it is completely model dependent and also if it plays a role in the retrieval of information from the collapsing system that forms the $BH$. In that regard, the factor $\pi/2$ in the ratio of entropy fluxes presented in Figure \ref{fig:thermalcontribution} is a noteworthy observation.

%

\section*{Acknowledgement}
I wish to thank Rodolfo Gambini, Jorge Pullin and Miguel Campiglia for valuable discussions and insights. This work was supported in part by ANII (FCE-1-2019-1-155865) and PEDECIBA.

\appendix
\section{Hawking radiation computation from Bogoliubov coefficients}\label{sec:Hawking}
\begin{figure}
\includegraphics[height=7cm]{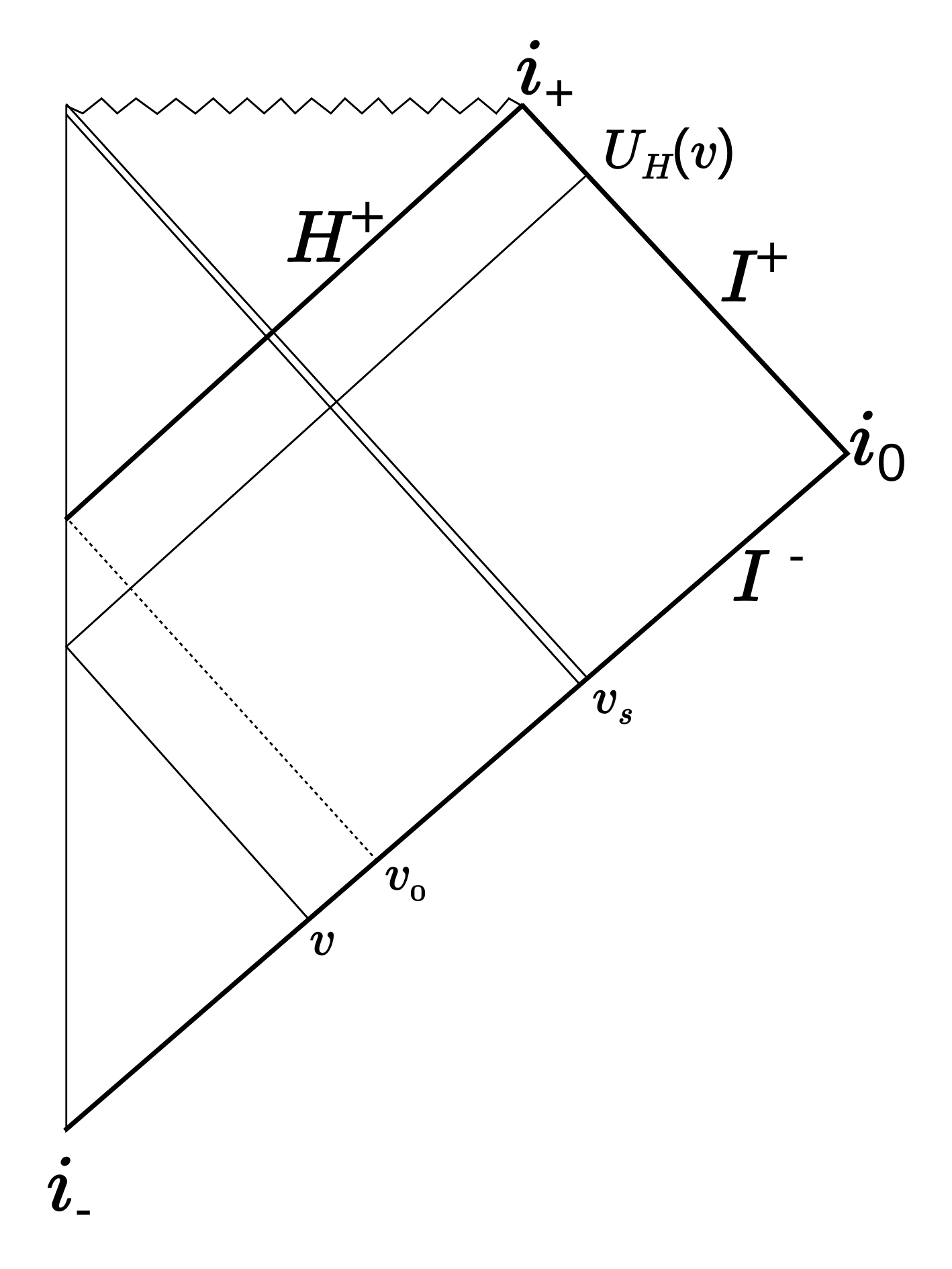}
\caption{Penrose-Carter diagram for metric (\ref{eq:metrica_shell}). $v_{s}$ indicate the positions at $I^-$ of the collapsing shell. Light rays sent in before $v_{0}$ reach $r=0$ and bounce back to $I^+$ at $U_H(v)$, whereas rays sent in after $v_{0}$ get trapped in the BH interior.}
\label{fig:espacio_tiempo_shell}
\end{figure}

In this appendix we reproduce known results for the thermal emission from a Vaidya space-time given by the metric
\begin{equation}
ds^{2}=-\left(1-\frac{2M(v)}{r}\right)dv^{2}+2dvdr+r^{2}d\Omega^{2},\label{eq:metrica_shell}
\end{equation}
with 
\begin{equation}
M(v)=M\theta(v-v_{s}),
\end{equation}
representing the collapse of a shell to form a BH of mass $M$ and whose Penrose-Carter diagram is shown is Figure \ref{fig:espacio_tiempo_shell}.
This analysis is the blueprint for the computation made in section \ref{sec:emision} and it provides the intuition to deduce the thermal spectrum in section \ref{sec:matriz_densidad}. In \citep{Eyheralde:2017jzd} we studied this model and described the transitory behavior modeling the beginning of Hawking radiation. Here we will focus in the better known result of late-time Hawking emission which, as Hawking argued \citep{h74}, is independent of the details of the collapse. To achieve this we quantize the massless scalar field in the near-horizon approximation as in section \ref{sec:emision}. Within this approximation, wave packet solutions of the K-G equation propagate through geometric optics, following the path of rays departing from $I^-$ at $v$ just before the formation of the horizon and arriving at $I^+$ at late-times $U_H(v)$. For this late-time modes the exact expression $U_H(v)$ can be approximated,
\begin{equation}
U_H(v)=v-4M\ln\left(\frac{v_{0}-v}{4M_{0}}\right)\approx v_{0}-4M\ln\left(\frac{v_{0}-v}{4M_{0}}\right).\label{eq:u(v)_Hawking}
\end{equation}
The construction of the \textit{in} quantization is straightforward because it only involves a basis of positive frequency modes at $I^-$ and the construction of the \textit{out-int} quantization is simple because the solution space can be split into positive frequency modes that do not register at $H^+$ (\textit{out} modes) and the ones that do not register at $I^+$ (int modes).
Hawking radiation if obtained computing the Bogoliubov coefficients (analogous to expression (\ref{eq:beta})) associated to the \textit{out} modes  and can be expressed in terms of an integral at $I^-$
\begin{equation}
\beta_{\omega\omega'}=-\frac{4M_{0}}{2\pi}\sqrt{\frac{\omega'}{\omega}}{\int_0^\infty}dxe^{-i\omega\left[v_{0}-4M\ln\left(x\right)\right]- i\omega'\left(v_{0}-4M_{0}x\right)}\label{eq:coef_bogol_H},
\end{equation}
where the change of variable 
$x=\frac{v_{0}-v}{4M_{0}}$ is used. For discrete wave packets (\ref{eq:paquete_def}) in the limit $\epsilon<<\omega_j$ we get
\begin{equation}
\beta_{\omega_j\omega'}=-\frac{4M_{0}}{2\pi}\sqrt{\frac{\omega'\epsilon}{\omega_j}}{\int_0^\infty}dxe^{i\omega_j\left[u_n-v_{0}+4M\ln\left(x\right)\right]}\textrm{sinc}\left(\left[u_n-v_{0}+4M\ln\left(x\right)\right]\frac{\epsilon}{2}\right)e^{- i\omega'\left(v_{0}-4M_{0}x\right)}\label{eq:coef_bogol_H_paquetes}.
\end{equation}
Now, from the Bogoliubov coefficients we can calculate the density matrix
\begin{equation}
\rho^{H}_{\omega_j\omega_k}={\int}_0^\infty d\omega'\beta_{\omega_j\omega'}\beta_{\omega_k\omega'}^{*},
\end{equation}
and in particular the expectation value of the number of particles per unit frequency detected at $I^+$
\begin{equation}
N^{H}_{\omega_j}=\rho^{H}_{\omega_j\omega_j},
\end{equation}
where we added the superscript ``$H$'' to indicate this is the calculation originally carried out by Hawking. In order to perform the $\omega'$ integration we can make a Wick-rotation in the integral (\ref{eq:coef_bogol_H_paquetes}) because the integrand is analytic in the complex plane. We get
\begin{equation}
\beta_{\omega_j\omega'}=-\frac{4M_{0}ie^{- i\omega'v_{0}}}{2\pi}\sqrt{\frac{\omega'\epsilon}{\omega_j}}{\int_0^\infty}dte^{i\omega_j\left[u_n-v_{0}+4M\ln\left(t\right)+i2M\pi\right]}\textrm{sinc}\left(\left[\frac{u_n-v_{0}}{4M}+\ln\left(t\right)+\frac{\pi}{2}i\right]2M\epsilon\right)e^{-\omega'4M_{0}t}
\end{equation}
and from here
{\footnotesize\begin{equation}
\rho^{H}_{\omega_j\omega_k}=\frac{e^{-2M(\omega_j+\omega_k)\pi}}{(2\pi)^2}\frac{\epsilon}{\sqrt{\omega_j\omega_k}}\stackrel[0]{+\infty}{\iint}\frac{dtd\bar{t}e^{i4M(\omega_j\ln(t)-\omega_k\ln(\bar{t}))}}{(t+\bar{t})^2}\textrm{sinc}\left(\left[\frac{u_n-v_{0}}{4M}+\ln\left(t\right)+\frac{\pi}{2}i\right]2M\epsilon\right)\textrm{sinc}\left(\left[\frac{u_n-v_{0}}{4M}+\ln\left(\bar{t}\right)-\frac{\pi}{2}i\right]2M\epsilon\right).
\end{equation}}This is a well known expression, but written as a double integral the final stationary thermal spectrum is not apparent upon first inspection. The process of computing both integrals is instructive to see where the different aspects of the thermal spectrum are codded in the integral expression. It also gives insight on how to handle more cumbersome examples.

As a first step, the change of variable $y=\ln\left(t\right)$ turns the integral into
{\footnotesize
\begin{equation}
\rho^{H}_{\omega_j\omega_k}=\frac{e^{-4M\bar{\omega}\pi}}{(4\pi)^2}\frac{\epsilon}{\sqrt{\omega_j\omega_k}}\stackrel[-\infty]{+\infty}{\iint}\frac{dyd\bar{y}e^{i4M\bar{\omega}(y-\bar{y})}e^{-i2M\Delta\omega(y+\bar{y})}}{\left[\textrm{cosh}\left(\frac{y-\bar{y}}{2}\right)\right]^2}\textrm{sinc}\left(\left[\frac{u_n-v_{0}}{4M}+y+\frac{\pi}{2}i\right]2M\epsilon\right)\textrm{sinc}\left(\left[\frac{u_n-v_{0}}{4M}+\bar{y}-\frac{\pi}{2}i\right]2M\epsilon\right),\label{eq:matriz_densidad_H}
\end{equation}}with $\bar{\omega}=\frac{\omega_j+\omega_k}{2}$ and $\Delta\omega=\omega_k-\omega_j$. A further change of variable $s=\frac{y-\bar{y}}{2}, \bar{s}=\frac{u_n-v_{0}}{4M}+\frac{y+\bar{y}}{2}$ allows to perform the $\bar{s}$ integration so
\begin{eqnarray}
\rho^{H}_{\omega_j\omega_k}&=&\frac{e^{-4M\bar{\omega}\pi}\epsilon e^{i\Delta\omega(u_n-v_0)}}{8\pi^2\sqrt{\omega_j\omega_k}}{\int_{-\infty}^\infty}\frac{dse^{i8M\bar{\omega}s}}{\left[\textrm{cosh}(s)\right]^2}\int_{-\infty}^\infty d\bar{s}e^{-i4M\Delta\omega\bar{s}}\textrm{sinc}\left(\left[\bar{s}+s+\frac{\pi}{2}i\right]2M\epsilon\right)\textrm{sinc}\left(\left[\bar{s}-s-\frac{\pi}{2}i\right]2M\epsilon\right)\nonumber\\
&=&\frac{e^{-4M\omega_j\pi}}{16M\omega_j\pi}{\int_{-\infty}^\infty}dse^{i8M\omega_js}\left[\sech(s)\right]^2\textrm{sinc}\left(\left[s+\frac{\pi}{2}i\right]4M\epsilon\right)\delta_{jk}.\label{eq:matriz_densidad_H_s}
\end{eqnarray}
The last step is obtained as an application of dual convolution theorem because the integration in $\bar{s}$ is the Fourier transform of a product and the Fourier transform of a $\rm{sinc}$ function is a unit pulse. The Kronecker $\delta$ makes apparent the lack of correlations between different frequencies. The constant factor $e^{-4M\pi\omega_j}$ hints to the thermal spectrum but a $1/\sinh(4M\pi\omega_j)$ is required from the last integral. A possible technique to compute it, that we haven't found in the literature, is a contour integration in the upper-half of the complex plane. The integral reduces to a infinite series of residues for second order poles at $z_n=(2n+1)\frac{\pi}{2}i$. In the limit $\epsilon<<\omega_j$ it simplifies to the infinite series
\begin{equation}
N^{H}_{\omega_j} =\frac{e^{-4M\omega_j\pi}}{16M\omega_j\pi}\sum_{n=0}^{\infty}2\pi i (-i8M\omega_j)e^{-4M\omega_j(2n+1)\pi}=\frac{1}{e^{8M\omega_j\pi}-1}.
\end{equation}
The luminosity ($L_H$) can be computed from the previous expression adding the contribution of all $\omega_j$ modes. However, to do the explicit computation it is convenient to go a step backward and do it from (\ref{eq:matriz_densidad_H_s}). We get
\begin{eqnarray}
L_H\frac{2\pi}{\epsilon}&=&\sum_j\hbar\omega_j N^{H}_{\omega_j}=\frac{\hbar}{16M\pi}{\int_{-\infty}^\infty}ds\left[\sech(s)\right]^2\sum_{j}e^{-4M\omega_j\pi+i8M\omega_js}\textrm{sinc}\left(\left[s+\frac{\pi}{2}i\right]4M\epsilon\right)=\\
&=&\frac{2\pi}{\epsilon}\frac{\hbar}{128 M^2\pi^2}{\int_{-\infty}^\infty}\frac{ds\left[\sech(s)\right]^2}{-2i\left(s+\frac{\pi}{2}i\right)},
\end{eqnarray}
where the explicit expression $\omega_j=(j+1/2)\epsilon$ is used to perform the summation in $j$. This integral can also be computed by a contour integration. Using the parity of $\sech$ and the fact that $\sech(z\pm i\pi)=-\sech(z)$ is can be rewritten as
\begin{eqnarray}
\stackrel[-\infty]{+\infty}{\int}\frac{ds\left[\sech(s)\right]^2}{-2i\left(s+\frac{\pi}{2}i\right)}&=&\frac{1}{2}\left[\stackrel[-\infty]{+\infty}{\int}\frac{ds\left[\sech(s)\right]^2}{-2i\left(s+\frac{\pi}{2}i\right)}+\stackrel[-\infty]{+\infty}{\int}\frac{ds\left[\sech(s)\right]^2}{-2i\left(-s+\frac{\pi}{2}i\right)}\right]=\frac{i}{4}\left[\stackrel[-\infty]{+\infty}{\int}\frac{ds\left[\sech(s)\right]^2}{s+\frac{\pi}{2}i}-\stackrel[-\infty]{+\infty}{\int}\frac{ds\left[\sech(s-i\pi)\right]^2}{s-\frac{\pi}{2}i}\right]\\
&=&\frac{i}{4}\oint_{\mathcal{C}} dz\frac{\left[\sech(z)\right]^2}{z+\frac{\pi}{2}i}=\frac{\pi}{6}
\end{eqnarray}
where $\mathcal{C}=\left\lbrace z:z\in\mathbb{R}\mid z\in\infty+i(0,-\pi)\mid z\in\mathbb{R}-i\pi\mid z\in-\infty+i(-\pi,0)\right\rbrace$ and the result is obtained by computing the residue of the third order pole at $z=-i\pi/2$. Finally we get the known result
\begin{equation}
L_H=\frac{\hbar}{768\pi M^2}
\end{equation}
which gives the contribution to the luminosity of all the s-modes in the near-horizon approximation.

\section{K-G equation in the near-horizon approximation}\label{sec:KG}

The first step to develop a quantum description of a field (in our case the free K-G field) is the study of the solution space in terms of data on a Cauchy surface. In the case of the K-G equation (\ref{eq:KG}) for the exterior (regions $I$ to $V$ in Figure \ref{fig:sandwich}) of the BH metric (\ref{eq:metrica}) some care is required due to the Cauchy horizon $C$. Regions $I$ to $IV$ together form a globally hyperbolic region of space-time and therefore data on any Cauchy surface fixes a solution in the bulk. $I^-$ is such a surface and this allows to build a quantization of the free scalar field using positive frequency spherical waves in that asymptotic region. One would also like to consider the null surface $\Sigma\equiv H^+\cup C \cup \left\lbrace I^+|u\leq v_{s2}\right\rbrace$. However $H^+$ and $C$ meet at the evaporation point $i_e$ where the centrifugal potential is singular so the initial value problem is ill defined there.
Despite the previous observation, in this paper we are using the near-horizon approximation to describe how solutions propagate and in this approximation the centrifugal potential is neglected. This allows to identify continuous data in $\Sigma$ with smooth solutions in the bulk.

A second problem related to the Cauchy horizon is the matching of solutions in regions $IV$ and $V$. We adopt the criteria that solutions should be smooth across $C$ with the possible exception of shock waves coming from the evaporation point $i_e$. The rational for this criteria is that regions $IV$ and $V$ are part of the same Minkowski exterior region with no sources for the field other than $i_e$. As we will see in what follows, this allow us to use $I^+\cup H^+$ instead of $\Sigma$ as a Cauchy surface to define an \textit{out-int} quantization of the field, in analogy to the construction presented in the previous appendix for a non evaporating BH.

 The previous condition allow us to study the propagation of spherical waves in the exterior of the BH provided two other usual conditions. One is the regularity at $r=0$ and guarantees the lack of source at the coordinates origin for the spherical waves. The second condition is the continuity of the wave fronts at the shells in Schwarzschild coordinates, which is the correct matching condition for null-shells \citep{Israel:1966rt}. Lets discuss the propagation of the relevant \textit{out},\textit{int} and \textit{in} modes.
 
\subsection{Spherical modes}
To study the propagation of waves in geometry (\ref{eq:metrica}) lets consider spherical waves of the form
 \begin{equation}
\chi_{lm}(r,v,\theta,\phi)=\frac{\phi_l(v,r)}{r}Y_{l}^m(\theta,\phi).
\end{equation}
For this kind of solutions, the K-G equation (\ref{eq:KG}) in the near-horizon approximation reduces to

\begin{equation}
\partial_u\partial_v\phi_l(v,r)=0,
\end{equation}
where the second null coordinate is $u=v-2r$ for the Minkowski regions $I,IV \& V$ and $u=v-2r-4M\ln\left|\frac{r-2M}{2M_0}\right|$ for regions $II \& III$. The equation becomes independent of $l$ and the general solution is simply
\begin{equation}
\phi(v,u)=F(v)+G(u).\label{eq:separacion}
\end{equation}
This equation is exact in the case of spherically symmetric modes ($l=0$) in regions $I,IV\&V$. For this reason from now on we reduce our analysis to such modes.
\subsection{Propagation of \textit{out} modes}
Our main interest is focused in the propagation of \textit{out} modes given by
\begin{eqnarray}
\phi^{out}(u)&\sim&\omega^{-1/2}e^{-i\omega u}, \quad at\quad I^+\\
\phi^{out}(v)&=&0, \quad at\quad H^+.
\end{eqnarray}
In region $V$ the general solution (\ref{eq:separacion}) imply $G(u)=\omega^{-1/2}e^{-i\omega u}$ and the regularity at $r=0$ imply $F(v)=-\omega^{-1/2}e^{-i\omega v}$ so the solution takes the form
\begin{equation}
\phi^{out}_{V}(v,u)=\omega^{-1/2}\left[e^{-i\omega u}-e^{-i\omega v}\right].
\end{equation}
The fact that this solutions vanish at $i_e$ allows to extended them continuously through $C$ and into region $IV$ as
$$\phi^{out}_{IV}(v,u)=\omega^{-1/2}\left[e^{-i\omega u}-e^{-i\omega v}\right].$$
Continuity at $v=v_{s2}$ and the value of the field at $H^+$ imply 
\begin{equation}
\phi^{out}_{II,III}(v,u)=\omega^{-1/2}\left[e^{-i\omega \hat{U}(u,v_{s2})}-e^{-i\omega v_{s2}}\right]
\end{equation}
in regions $II \& III$ where
\begin{equation}
\hat{U}(u,v)= v-4M-4M\textrm{W}\left[\textrm{sign}(r-2M)\exp\left(\frac{v-u-4M}{4M}\right)\frac{M_0}{M}\right].\label{eq:Uhat}
\end{equation}
Finally, due to the continuity at $v=v_{s1}$ and the regularity at $r=0$ the \textit{out} modes becomes
\begin{eqnarray}
\phi^{out}_{I}&=&\omega^{-1/2}\left[e^{-i\omega U(u)}-e^{-i\omega v_{s2}}\right] \quad for\quad v>v_{02}\\
\phi^{out}_{I}&=&\omega^{-1/2}\left[e^{-i\omega U(u)}-e^{-i\omega U(v)}\right], \quad for\quad v<v_{02}
\end{eqnarray}
in region $I$ with $U(v)$ given by (\ref{eq:u(v)}). Notice that the restriction to $\Sigma$ gives a continuous function. These are trivially continuous across $IV$ and $V$ and also, when restricted to $H^+\cup C$, we get the continuous data
\begin{equation}
\chi^{out}(r,v)|_{H^+\cup C}=\frac{1}{4\pi r\sqrt{\omega}}\times\left\{ \begin{array}{lcl}
e^{-i\omega v_{s2}}-e^{-i\omega v},\quad v>v_{s2}\\
0,\quad\qquad\qquad\qquad v<v_{s2}
\end{array}\right.\label{eq:out_sigma}
\end{equation}
Finally, if the modes are restricted to a null surface of constant\footnote{One way to arrive at $I^-$, parameterized by $v$ and angular coordinates, is to take the limit $u_0\to\ -\infty$ for fixed $v$.} $u=u_0$ we get

\begin{equation}
\chi^{out}(r,v)|_{u=u_0}=\frac{1}{4\pi r\sqrt{\omega}}\left[e^{-i\omega u_0}-e^{-i\omega v_{s2}}\right]+\frac{1}{4\pi r\sqrt{\omega}}\times\left\{ \begin{array}{lcl}
e^{-i\omega v_{s2}}-e^{-i\omega v},\qquad v>v_{s2}\\
0,\qquad\qquad\qquad\qquad v_{s2}>v>v_{02}\\
e^{-i\omega v_{s2}}-e^{-i\omega U(v)},\qquad v<v_{02}\\
\end{array}\right.\label{eq:out_Imenos}
\end{equation}
This represents radiation registering on $I^-$ only at regions $I$ and $IV$.

\subsection{Propagation of \textit{int} modes}

The \textit{int} modes are generated by data at the horizon and no radiation at $I^+$. Given
\begin{eqnarray}
\phi^{int}(u)&=&0 \quad at\quad I^+\\
\phi^{int}(v)&=&F(v), \quad at\quad H^+,
\end{eqnarray}
where $F$ is a smooth function with the condition $F(v_{02})=F(v_{s2})=0$ in order to avoid sources at $r=0$ and at $i_e$. In the near-horizon approximation this implies
\begin{eqnarray}
\phi^{int}_{I}(v,u)&=&0 \quad for\quad v<v_{02}\\
\phi^{int}_{I}(v,u)&=&F(v), \quad for \quad v>v_{02}
\end{eqnarray}
in region $I$ and $\phi^{int}_{II}(v,u)=F(v)$ in region $II$. Finally $\phi^{int}_{IV,V}=0$ in the remaining Minkowski regions $IV \& V$ where no radiation is present. The data in $H^+\cup C$ is obviously continuous. In fact when restricted to $\Sigma$ this modes take the value
\begin{equation}
\chi^{int}|_{\Sigma}=\frac{1}{4\pi r\sqrt{\omega}}\times\left\{ \begin{array}{lcl}
0,\quad\qquad u <v_{s2}\quad at\quad I^+\\
0,\quad\qquad v>v_{s2}\quad at\quad C\\
F(v),\qquad v_{s2}>v>v_{02}\\
\end{array}\right.\label{eq:int_sigma}
\end{equation}
This is valid for any surface of $u=u_0$ to the past and towards $I^-$ where we simply get
\begin{equation}
\chi^{int}(r,v)=\frac{1}{4\pi r\sqrt{\omega}}\times\left\{ \begin{array}{lcl}
0,\qquad v>v_{s2}\\
F(v),\quad v_{s2}>v>v_{02}\\
0,\qquad v<v_{02}\\
\end{array}\right.\label{eq:int_Imenos}
\end{equation}

\subsection{Propagation of \textit{in} modes}

Starting with spherical modes in the past, given by
\begin{equation}
\phi^{in}(v)\sim\omega'^{-1/2}e^{-i\omega' v}\quad at \quad I^-,
\end{equation}
the regularity condition at $r=0$ imply
\begin{equation}
\phi^{in}_{I}(v,u)=\omega'^{-1/2}\left[e^{-i\omega' v}-e^{-i\omega' u}\right].
\end{equation}
Due to the continuity at $v=v_{s1}$ and the near-horizon approximation, in region $II$ and $III$ they become 
\begin{equation}
\phi^{in}_{II,III}(v,u)=\omega'^{-1/2}\left[e^{-i\omega' v}-e^{-i\omega' \hat{U}(u,v_{s1})}\right].
\end{equation}
Then, continuity at $v=v_{s2}$ imply
\begin{equation}
\phi^{in}_{IV}=\omega'^{-1/2}\left[e^{-i\omega' v}-e^{-i\omega' \tilde{U}(u)}\right],
\end{equation}
where
\begin{equation}
\tilde{U}(u)=v_{s1}-4M-4M\textrm{W}\left[\frac{v_{s2}-u-4M}{4M}\exp\left(\frac{v_{s1}-u-4M}{4M}\right)\right]\label{eq:Utilde}.
\end{equation} 
If we restrict the \textit{int} modes to the Cauchy surface $\Sigma$ we find a continuous data as expected, however $\phi_{IV}$ can not be continuously extended to region $V$ from data in $C$ because a solution to the free K-G equation has to vanish at $r=0$. A natural extension would be to match the solutions across $C$ up to a shock wave coming from $i_e$. That gives us

\begin{equation}
\phi^{in}_{V}=\omega'^{-1/2}\left[e^{-i\omega' v}-e^{-i\omega' u}\right].
\end{equation}
and the shock wave at $C$ is fixed to
\begin{equation}
\phi^{in}_{V}(v,v_{s2})-\phi^{in}_{IV}(v,v_{s2})= \omega'^{-1/2}\left[e^{-i\omega' v_{02}}-e^{-i\omega' v_{s2}}\right].
\end{equation}
This extension, however natural, doesn't come with a natural extension of the K-G product to $I^+$. 

\subsection{Cauchy surfaces and the Klein-Gordon product}
We have shown that \textit{out} and \textit{int} modes span radiative solutions that register in disjoint regions of $\Sigma$ (see expressions \ref{eq:out_sigma} and \ref{eq:int_sigma}). Therefore they are orthogonal with respect to the K-G product. In the near-horizon approximation they also register in different regions of $I^-$ as can be seen explicitly from  (\ref{eq:out_Imenos}) and (\ref{eq:int_Imenos}). An important property of this modes is that the K-G product is invariant when $\Sigma$ is replaced by $H^+\cup I^+$. For \textit{int} modes this is trivial because they do not register at $C$ but for modes that have data $G(v)$ at $C$, the K-G product is invariant only if the field in region $V$ has the form $\phi(v,u)=G(v)-G(u)$. \textit{out} modes fulfill this condition automatically, but any mode with data $G(v_{s2})\neq 0$ at $i_e$ has to present a discontinuity at $C$. This is what happens with general solutions spanned by \textit{in} modes. We can identify this discontinuity as a shock wave at $C$ on top of the usual radiative modes but this doesn't give as a prescription for the extension of the K-G product to the future of $C$. Nevertheless we can compute the K-G product at $\Sigma$ projecting back the \textit{out} modes from $H^+\cup I^+$ in an unambiguous way and this is all we need to compute the relevant Bogoliubov coefficients.


\end{document}